\documentclass[preprint, a4paper, 12pt, aps, prb, amsmath]{revtex4-1}
\usepackage[utf8]{inputenc}
\usepackage{graphicx}
\usepackage{subfigure}
\usepackage{caption}

\newcommand{\Ref}[1]{[\onlinecite{#1}]}

\newcommand{\Secref}[1]{Section \ref{sec:#1}}
\newcommand{\Eqnref}[1]{Eq.~\eqref{eq:#1}}
\newcommand{\Eqsref}[1]{Eqs.~\eqref{eq:#1}}
\newcommand{\Figref}[1]{Fig.~\ref{fig:#1}}
\newcommand{\Figsref}[1]{Figs.~\ref{fig:#1}}
\newcommand{\bigexp}[1]{\ensuremath{\exp \left( #1 \right)}}
\newcommand{\taud}{\ensuremath{\tau_{\mathrm{d}}}}
\newcommand{\mean}[1]{\ensuremath{\langle #1 \rangle}}
\newcommand{\tauw}{\ensuremath{\tau_{\mathrm{w}}}}
\newcommand{\var}[1]{\ensuremath{\text{var}\left( #1 \right)}}
\newcommand{\skw}[1]{\ensuremath{S\left(#1\right)}}
\newcommand{\krt}[1]{\ensuremath{F\left(#1\right)}}
\newcommand{\tauwd}{\ensuremath{\frac{\tau_\mathrm{w}}{\tau_\mathrm{d}}}}
\newcommand{\taudw}{\ensuremath{\frac{\tau_\mathrm{d}}{\tau_\mathrm{w}}}}
\newcommand{\msemu}{\ensuremath{\mathrm{MSE} ( \widehat{\mu} )}}
\newcommand{\msevar}{\ensuremath{\mathrm{MSE} ( \widehat{\sigma^2} )}}
\newcommand{\mse}[1]{\ensuremath{\mathrm{MSE} ( #1 )}}
\newcommand{\estmu}{\ensuremath{\widehat{\mu}}}
\newcommand{\estvar}{\ensuremath{\widehat{\sigma^2}}}
\newcommand{\estskw}{\ensuremath{\widehat{S}}}
\newcommand{\estkrt}{\ensuremath{\widehat{F}}}
\newcommand{\estskwG}{\ensuremath{\widehat{S}_\Gamma}}
\newcommand{\estskwGm}{\ensuremath{\widehat{S}_{\Gamma, m}}}
\newcommand{\estkrtGm}{\ensuremath{\widehat{F}_{\Gamma, m}}}
\newcommand{\estkrtG}{\ensuremath{\widehat{F}_\Gamma}}
\newcommand{\estskwm}{\ensuremath{\widehat{S}_m}}
\newcommand{\estkrtm}{\ensuremath{\widehat{F}_m}}
\newcommand{\est}[1]{\ensuremath{\widehat{ #1 }}}
\newcommand{\textvar}[1]{\ensuremath{\text{var}(#1)}}
\newcommand{\covmusigma}{\ensuremath{\mathrm{COV} (\widehat{\mu}, \widehat{\sigma^2} )}}

\begin{document}
\title{Convergence of statistical moments of particle density time series in scrape-off layer plasmas}
\author{R.~Kube}
\email[E-mail:]{ralph.kube@uit.no}

\author{O.~E.~Garcia}
\affiliation{Department of Physics and Technology,
UiT - The Arctic University of Norway,
N-9037 Tromsø, Norway}

\date{\today}
\begin{abstract}
Particle density fluctuations in the scrape-off layer of magnetically confined
plasmas, as measured by gas-puff imaging or Langmuir probes, are modeled as the 
realization of a stochastic process in which a superposition of pulses with a
fixed shape, an exponential distribution of waiting times and amplitudes represents
the radial motion of blob-like structures. With an analytic formulation of the
process at hand, we derive expressions for the mean squared error on estimators of
sample mean and sample variance as a function of sample length, sampling frequency,
and the parameters of the stochastic process.
Employing that the probability distribution function of a particularly relevant 
stochastic process is given by the gamma distribution, we derive estimators for sample 
skewness and kurtosis, and expressions for the mean squared error on these 
estimators. 
Numerically generated synthetic time series are used to verify the proposed estimators, 
the sample length dependency of their mean squared errors, and their performance. 
We find that estimators for sample skewness and kurtosis based on the gamma 
distribution are more precise and more accurate than common estimators based on the method of moments.
\end{abstract}
\maketitle
\section{Introduction}
Turbulent transport in the edge of magnetically confined plasmas is a key issue to be 
understood on the way to improved plasma confinement, and ultimately commercially
viable fusion power.
Within the last-closed magnetic flux surface, time series of the particle density present 
small relative fluctuation amplitudes and Gaussian amplitude statistics. The picture 
in the scrape-off layer (SOL) is quite different. Time series of the particle density, 
as obtained by single point measurements, present a relative fluctuation level of order 
unity. Sample coefficients of skewness and excess kurtosis
\cite{foot-1} 
 of these time series are non 
vanishing and sample histograms feature elevated tails. This implies that the
deviation from normality is caused by the frequent occurrence of large amplitude events
\Ref{antar-2001, antar-2003, xu-2005, agostini-2007, dewhurst-2008}. 

These features of fluctuations in the scrape-off layer are attributed to the radially 
outwards motion of large amplitude plasma filaments, or blobs.
Time series of the plasma particle density obtained experimentally
\Ref{boedo-2003, agostini-2007, xu-2009, tanaka-2009, cheng-2010, nold-2010} 
and by numerical simulations
\Ref{russell-2007, garcia-2007-tcv, russell-2007, myra-2008, militello-2012} 
show that estimated coefficients of skewness and excess kurtosis \Ref{balanda-1988} increase 
radially outwards with distance to the last closed flux surface. At the same time one observes 
a parabolic relationship between these two coefficients 
and that the coefficient of skewness vanishes close to the last closed flux surface
\Ref{russell-2007, graves-2005, sattin-2006, labit-2007, cheng-2010, garcia-2013}. 
%Analysis of large data sets of particle density time series in TORPEX and
%TCV showed, that this relationship is due to drift-interchange
%turbulence and well described by a beta distribution 
%\Ref{labit-2007, labit-2007_PPCF}, which collapses to a gamma distribution
%in the case of positive skewness (The beta-distribution allows for
%negative values of skewness, while $\mathrm{S}>0$ for a gamma). \\

% Move the two paragaphs below to discussion
%Only few attempts have been made to present a model for the observed
%non-Gaussian statistics in the scrape-off layer. One model propagates
%normal distributed potential measurements on to a PDF for the
%density using a vorticity-free blob model \cite{sattin-2004} so that 
%the resulting PDF is similar to a log-normal distribution.
%
%Other attempts at modeling the probability distribution function of
%the particle flux in the scrape-off layer find a Gumbel distribution
%as a result of the non-linearities in reduced Braginskii models
%\Ref{anderson-2010}.
%
Recently, it was proposed to model the observed particle density time series by a shot 
noise process \Ref{rice-1977}, that is, a random superposition of pulses corresponding to blob structures
propagating through the scrape-off layer \Ref{garcia-2012}.
Describing individual pulses by an exponentially decaying waveform with exponentially 
distributed pulse amplitudes and waiting time between consecutive pulses leads to 
a Gamma distribution for the particle density amplitudes \Ref{garcia-2012, garcia-2006-tcv}.
In this model, the shape and scale parameter of the resulting Gamma distribution can be 
expressed by the pulse duration time and average pulse waiting time.

In order to compare predictions from this stochastic model to experimental
measurements, long time series are needed, as to calculate statistical
averages with high accuracy. Due to a finite correlation time of the
fluctuations, an increased sampling frequency may increase the number of 
statistically independent samples only up to a certain fraction. Then,
only an increase in the length of the time series may increase the number
of independent samples. This poses a problem for Langmuir probes, which are
subject to large heat fluxes and may therefore only be dwelled in the scrape-off 
layer for a limited amount of time. Optical diagnostics on the other hand, may 
observe for an extended time interval but have other drawbacks, as 
for example the need to inject a neutral gas into the plasma to increase the
signal to noise ratio, and that the signal intensity depends sensitively on 
the plasma parameters \Ref{zweben-2002, stotler-2003, cziegler-phd}. 

This work builds on the stochastic model presented in Ref.~\Ref{garcia-2012}
by proposing estimators for the mean, variance, skewness and excess kurtosis 
of a shot noise process and deriving expressions of their mean squared error as a function of 
sample length, sampling frequency, pulse amplitude, and duration, and waiting time.
Subsequently, we generate synthetic time series of the shot noise
process at hand. The mean squared error of the proposed estimators
is computed of these time series and their dependence on the sampling parameters and the process
parameters is discussed. 

This paper is organized as follows. \Secref{sol_turbulence} introduces the stochastic 
process that models particle density fluctuations and the correlation function of this 
process. 
In \Secref{estimators} we propose statistical estimators to be used for the shot-noise
process and derive expressions for the mean squared error on these estimators.
A comparison of the introduced estimators and expressions for their mean squared
error to results from analysis of synthetic time series of a shot noise process is
given in \Secref{synth_ts}.
A summary and conclusions are given in \Secref{conclusion}.

%
%%%%%%%%%%%%%%%%%%%%%%%%%%%%%%%%%%%%%%%%%%%%%%%%%%%%%%%%%%%%%%%%%%%%%%%%%%%%%%%%%%%%%%%%%%%
%
%%%%%%%%%%%%%%%%%%%%%%%%%%%%%%%%%%%%%%%%%%%%%%%%%%%%%%%%%%%%%%%%%%%%%%%%%%%%%%%%%%%%%%%%%%%
%
\section{Stochastic model}
\label{sec:sol_turbulence}
A stochastic process formed by superposing the realization of independent
random events is commonly called a shot noise process \Ref{rice-1977, pecseli-book-fluct}. 
Denoting the pulse form as $\psi(t)$, the amplitude as $A_k$, and the arrival time as 
$t_k$, a realization of a shot noise process with $K$ pulses is written as
%92G
\begin{align}
    \Phi_K(t) & = \sum\limits_{k=1}^{K} A_k \psi(t - t_k). \label{eq:def_shotnoise}
\end{align}
To model particle density time series in the scrape-off layer by a stochastic
process, the salient features of experimental measurements have to be reproduced
by it.

Analysis of experimental measurement data from tokamak plasmas, 
\Ref{antar-2003, dewhurst-2008, xu-2005, boedo-2003, cheng-2010, garcia-2007-tcv, garcia-2006-tcv}
as well as numerical simulations 
\Ref{garcia-2006-tcv, bian-2003, garcia-2006, kube-2011},
have revealed large
amplitude bursts with an asymmetric wave form, featuring a fast rise time and
a slow exponential decay. The burst duration is found to be independent of the
burst amplitude and the plasma parameters in the scrape-off layer \Ref{garcia-2009, garcia-2013}.
The waveform to be used in \Eqnref{def_shotnoise} is thus modeled as
\begin{align}
    \psi_k(t) = \bigexp{-\frac{t}{\taud}} \Theta(t), \label{eq:pulse_shape_exp}
\end{align}
where $\taud$ is the pulse duration time and $\Theta$ denotes the Heaviside step function. 
Analysis of long data time series further reveals that the pulse amplitudes $A$
are exponentially distributed \Ref{garcia-2013},
\begin{align}
    P_A(A) = \frac{1}{\mean{A}}\bigexp{-\frac{A}{\mean{A}}}. \label{eq:dist_A}
\end{align}
Here $\mean{A}$ is the scale parameter of the exponential distribution, and
$\mean{\cdot}$ denotes an ensemble average.
The waiting times between consecutive bursts are found to be exponentially 
distributed \Ref{antar-2001, antar-2003, garcia-2013, furchert-2013}.
Postulating uniformly distributed pulse arrival times $t$ on an interval
length $T$, $P_t(t) = 1/T$, it follows that the total number of pulses in a fixed 
time interval, $K$, is Poisson distributed and that the waiting time between
consecutive pulses, $\tauw$, is therefore 
also exponentially distributed \Ref{pecseli-book-fluct}.

Under these assumptions it was shown that the stationary amplitude distribution of 
the stochastic process given by \Eqnref{def_shotnoise} is a Gamma distribution 
\Ref{garcia-2012},
\begin{align}
    P_\Phi(\Phi) & = \frac{1}{\Gamma(\gamma)} \left(\frac{\gamma}{\mean{\Phi}}\right)^{\gamma} \Phi^{\gamma-1} \bigexp{-\frac{\gamma \Phi}{\mean{\Phi}}},
    \label{eq:gamma_PDF}
\end{align}
with the shape parameter given by the ratio of pulse duration time to the average
pulse waiting time
\begin{align}
    \gamma = \frac{\taud}{\tauw}. \label{eq:def_gamma_burst}
\end{align}
This ratio describes the intermittency of the shot noise process.
In the limit $\gamma \ll 1$, individual pulses appear isolated whereas $\gamma \gg 1$ 
describes the case of strong pulse overlap. In Ref. \Ref{garcia-2012} it was further 
shown that the mean, $\mean{\Phi}$, the
variance, $\var{\Phi} = \mean{\left( \Phi - \mean{\Phi}\right)^2}$, 
the coefficient of skewness, $\skw{\Phi}$, 
and the coefficient of flatness, or excess kurtosis, $\krt{\Phi}$, are in this case given by
\begin{subequations}
    \label{eq:sn_mom}
    \begin{align}
        \mean{\Phi} & = \mean{A} \frac{\taud}{\tauw}, & \var{\Phi} & = \mean{A}^2 \frac{\taud}{\tauw} \label{eq:sn_mom_mean_sigma},\\
        \skw{\Phi}  & = 2\left( \tauwd \right)^{1/2}, & \krt{\Phi} & = 6 \tauwd. \label{eq:sn_mom_SK}
    \end{align}
\end{subequations}
Thus, the parameters of the shot noise process, $\taud / \tauw$, and $\mean{A}$,
may be estimated from the two lowest order moments of a time series. 
Before we proceed in the next section to define estimators for these 
quantities, we continue by deriving 
an expression for the correlation function of the signal given by \Eqnref{def_shotnoise}. 
Formally, we follow the method outlined in Ref.~\Ref{pecseli-book-fluct}.

Given the definition of a correlation function, we average over the pulse arrival time
and amplitude distribution and use that for an exponentially distributed pulse
amplitude, $\mean{A^n} = n! \mean{A}$ holds. This gives
\begin{align}
\mean{\Phi_K(t) \Phi_K(t + \tau)}
    & = \int\limits_{0}^{T} \mathrm{d}t_1 P_t(t_1) \int\limits_{0}^{\infty} \mathrm{d}A_1 P_A(A_1) \cdots \int\limits_{0}^{T} \mathrm{d}t_K P_t(t_K) \int\limits_{0}^{\infty} \mathrm{d}A_K P_A(A_K) \times \nonumber \\
    & \hspace{4ex} \sum\limits_{p=1}^{K} \sum\limits_{q=1}^{K} A_p \psi(t - t_p)\, A_q \psi(t + \tau - t_q) \nonumber \\
    & = \mean{A^2} \sum\limits_{p = 1}^{K} \int\limits_{0}^{T} \frac{\mathrm{d}t_p}{T} \psi(t - t_p) \psi(t + \tau - t_p) \nonumber \\
    & + \mean{A}^2 \sum \limits_{p \neq q} \int\limits_{0}^{T} \frac{\mathrm{d}t_p}{T} \int\limits_{0}^{T} \frac{\mathrm{d}t_q}{T} \psi(t - t_p) \psi(t + \tau - t_q). 
    \label{eq:corr2_stop}
\end{align}
Here, we have divided the sum in two parts. The first part consists of $K$ terms
where $p = q$ and the second part consists of $K(K-1)$ terms where $p \neq q$.
The integral over a single pulse is given by
\begin{align}
    \int\limits_{0}^{T} \mathrm{d}t_p\; P_t(t_p) \psi(t - t_p)  
    %= \int\limits_{0}^{T} \frac{\mathrm{d}t_p}{T} \Theta(t - t_p) \bigexp{-\frac{t - t_p}{\taud}} 
    = \frac{\taud}{T} \left[ 1 - \bigexp{ -\frac{t}{\taud} }\right], \label{eq:integral_1pulse}
\end{align}
where the boundary term $\exp(-t / \taud)$ arises due to the 
finite integration domain. For observation times $t \gg \taud$ this term 
vanishes and in the following we neglect it by ignoring the initial transient part
of the time series where only few pulses contribute to the amplitude of the
signal. 

Within the same approximation, the integral of the product of two independent
pulses is given by
\begin{align*}
    \int\limits_{0}^{T} \mathrm{d}t_p\; P(t_p) \psi(t - t_p) \psi(t + \tau - t_p)
    = \frac{\taud}{2 T} \bigexp{-\frac{|\tau|}{\taud}}.
\end{align*}
Substituting these two results into \Eqnref{corr2_stop}, we average over the number of pulses 
occurring in $[0:T]$. Using that the total number of pulses is Poisson distributed and that 
the average waiting time between consecutive pulses is given by $\tauw = T / \mean{K}$, 
we evaluate the two-point correlation function of \Eqnref{def_shotnoise} as
\begin{align}
    \mean{\Phi(t) \Phi(t + \tau)} & = \mean{A}^2 \taudw \left[ \bigexp{-\frac{|\tau|}{\taud}} + \taudw \right]. \label{eq:sn_corr_twopoint}
\end{align}
Comparing this expression to the ensemble average of the model at hand,
\Eqnref{sn_mom_mean_sigma}, we find 
$\mean{\Phi(t) \Phi(t + \tau)} = \mean{\Phi(t)} \left[ \mean{A} \exp\left( -|\tau| / \taud \right) + \mean{\Phi(t)} \right].$
For $\tau \rightarrow \infty$, the correlation function decays exponentially
to the square of the ensemble average. 

\section{Statistical estimators for the Gamma distribution}
\label{sec:estimators}
The Gamma distribution is a continuous probability distribution with
a shape parameter $\gamma$ and a scale parameter $\theta$. The
probability distribution function (PDF) of a gamma distributed random
variable $X > 0$ is given by
\begin{align}
    P_X(X;\gamma, \theta) & = \frac{X^{\gamma-1}}{\theta^{\gamma} \Gamma(\gamma)} \bigexp{-\frac{X}{\theta}},
    \label{eq:def_Gamma}
\end{align}
where $\Gamma(x) = \int\limits_{0}^{\infty} \mathrm{d}u\,u^{x - 1}e^{-u}$  denotes the gamma function. 
%Its cumulative distribution function is given by
%%
%\begin{align}
%    F_X(X; \gamma, \theta) & = \frac{1}{\Gamma(\gamma)} \widetilde{\Gamma} \left(\gamma, \frac{X}{\theta}\right)
%    \label{eq:def_Gamma_cdf}
%\end{align}
%%
%where $\widetilde{\Gamma} (\gamma, X) = \int_{0}^{X} t^{\gamma -1}e^{-\gamma} \mathrm{d}t\;$ is
%the upper incomplete gamma function. 
Statistics of a random variable are often described in terms of the moments 
of its distribution function, which are defined as
\begin{align*}
    m_k = \int \limits_{0}^{\infty} \mathrm{d}X\; P_X(X; \gamma, \theta) x^k,
\end{align*}
and centered moments of its distribution function, defined as
\begin{align*}
    \mu_k = \int \limits_{0}^{\infty} \mathrm{d}X\; \left[P_X(X; \gamma, \theta) - m_1\right]^k.
\end{align*}
Common statistics used to describe a random variable are the mean $\mu = m_1$, the
variance $\sigma^2 = \mu_2$, skewness $S = \mu_3 / \mu_2^{3/2}$ and excess kurtosis,
or flatness,  $F= \mu_4 / \mu_2^2 - 3$. Skewness and excess kurtosis are well established 
measures to characterize asymmetry and elevated tails of a probability distribution function. 
%While skewness measures the asymmetry of the PDF, kurtosis has been described as the
%\emph{scale-free motion of probability density from its shoulders to
%its center and tails} \Ref{balanda-1988}. 
For a Gamma distribution, the moments relate to the shape and scale parameter as
\begin{align*}
    m_1 = \gamma \theta, \qquad \mu_2 = \gamma \theta^2, \qquad
    \mu_3 = 2 \gamma \theta^3, \qquad \mu_4 = 6 \gamma \theta^4,
\end{align*}
and coefficients of skewness and excess kurtosis are given in terms of the
shape parameter by
\begin{align*}
    S = \frac{\mu_3}{\mu_2^{3/2}} = \frac{2}{\sqrt{\gamma}},  \qquad \qquad
    F = \frac{\mu_4}{\mu_2^2} - 3 = \frac{6}{\gamma}. 
\end{align*}
For the process described by \Eqnref{def_shotnoise}, $\gamma$ is given by the ratio of 
pulse duration time to pulse waiting time, so that skewness and excess kurtosis assume large 
values in the case of strong intermittency, that is, weak pulse overlap.

In practice, a realization of a shot noise process, given by \Eqnref{def_shotnoise}, is typically
sampled  for a finite time $T$ at a constant sampling rate $1 / \triangle_t$ as to obtain 
a total of $N = T / \triangle_t$ samples. When a sample of the process is taken after the 
initial transient, where only few pulses contribute to the amplitude, the probability 
distribution function of the sampled amplitudes is given by the stationary distribution 
function of the process described by \Eqnref{gamma_PDF}.

%The most common method of estimating sample statistics is based on the methods
%of moments (citation to pearson?). For small sample sizes various estimators
%of skewness and kurtosis have been proposed, as to trade precision for
%accuracy. However, in the large sample limit, estimators for sample skewness
%and kurtosis show to all give identical results \Ref{joanes-1998}.
%In the case where the underlying distribution of a sample is known,
%maximum likelihood estimators are commonly employed to
%robustly estimate the parameters of the the distribution. Their drawback
%is that the equations are analytically untraceable and need to be
%solved numerically (citation: this one MLE paper and Minka for Gamma f.~ex.~)
%\Ref{minka-gamma}.
%
%When presented with a set of sampled data of a process where the underlying distribution 
%is known, maximum likelihood estimators are commonly employed to robustly estimate the 
%unknown parameters of the underlying distribution \Ref{cam-1990}.
%The drawback of this method is that the maximum likelihood equations for estimating the
%shape- and scale-parameter of a Gamma distribution are analytically untractable
%and have to be solved numerically \Ref{minka-gamma}.

We wish to estimate the moments of the distribution function 
underlying a set of $N$ data points, $\{x_i\}_{i=1}^{N}$, which are now taken to be samples of a 
continuous shot noise process, obtained at discrete sampling times 
$t_i = i \cdot \triangle_t$, $x_i = \Phi(t_i)$.
Using the method of moments, estimators of mean, variance, skewness,
and excess kurtosis are defined as
\begin{subequations}
    \begin{align}
        \widehat{\mu} & = \frac{1}{N} \sum\limits_{i=1}^{N} x_i,
        & \widehat{\sigma^2} & = \frac{1}{N-1} \sum \limits_{i=1}^{N} \left(x_i - \estmu \right)^2, \label{eq:est_mu_var} \\
        \est{S} & = \frac{\sum\limits_{i=1}^{N} \left(x_i - \estmu\right)^3}{\left(\sum \limits_{i=1}^{N} \left(x_i - \estmu\right)^2\right)^{3/2}}, 
        & \est{F} & = \frac{\sum\limits_{i=1}^{N} \left(x_i - \estmu\right)^4}{\left(\sum \limits_{i=1}^{N} \left(x_i - \estmu\right)^2\right)^2} - 3.\label{eq:est_SKmom}
    \end{align}
\end{subequations}
Here, and in the following, hatted quantities denote an estimator.
Building on these, we further define an estimator for the intermittency parameter
of the shot noise process according to \Eqnref{sn_mom_mean_sigma}
\begin{align}
    \est{\gamma} & = \frac{\estmu^2}{\estvar}. \label{eq:def_est_gamma}
\end{align}
We use this estimator to define alternative estimators for skewness and excess kurtosis as
\begin{align}
    \estskwG & = \frac{2}{\sqrt{\est{\gamma}}}, & \estkrtG & = \frac{6}{\est{\gamma}}. \label{eq:def_est_SG_KG}
\end{align}
in accordance with \Eqnref{sn_mom_SK}. 

In general, any estimator $\widehat{U}$ is a function of $N$ random variables and
therefore a random variable itself. A desired property of any estimator is that
with increasing argument sample size its value converges to the true value that one wishes
to estimate. The notion of distance to the true value is commonly measured by the
mean squared error on the estimator $\widehat{U}$, given by
\begin{align}
    % Definition of the mean-squqred error on an estimaror
    \mse{\est{U}}= \text{var} (\widehat{U}) + \mathrm{bias}(\widehat{U}, U)^2, \label{eq:def_MSE_est}
\end{align}
where $\textvar{ \widehat{U} } = \mean{(\widehat{U} - \mean{\widehat{U}})^2}$, 
$\mathrm{bias}(\widehat{U}, U) = \mean{\mean{\widehat{U}} - U}$,  and
$\langle \cdot \rangle$ denotes the ensemble average.
When \Eqnref{est_mu_var} is applied to a sample of $N$ normally distributed and uncorrelated
random variables, it can be shown that $\mathrm{bias}(\estmu, \mu) = 0$, 
$\mathrm{bias}(\estvar, \sigma^2) = 0$, and that the mean squared error of both estimators 
is inversely proportional to the sample size, $\mse{\estmu} \sim N^{-1}$, and 
$\mse{\estvar} \sim N^{-1}$.
For a sample of gamma distributed and independent random variables, 
$\mean{\estmu} = \mu = \gamma \theta$ and $\mean{\estvar} = \mu_2 = \gamma \theta^2$ holds.
Thus the estimators defined in \Eqnref{est_mu_var} have vanishing bias and their 
mean-square error is given by their respective variance, 
$\textvar{ \estmu }$ and $\textvar{ \estvar }$. 

%The rate of convergence of the estimators given in \Eqnref{est_mu_var}
%and \Eqnref{def_est_SG_KG}, to their asymptotic values as a function of
%sample length is thus given by their mean square error. 

%
%
%Due to the finite sample length $N$, the estimators defined above do not yield the exact 
%value of the analytic results, given in \Eqnref{sn_mean_sigma}. This deviation
%is quantified by the mean-square error of an estimator $\widehat{U}$,
%defined as
%%
%\begin{align}
%    % Definition of the mean-squqred error on an estimaror
%    \mse{\est{U}}= \var{\widehat{U}} + \mathrm{Bias}(\widehat{U}, U)^2. \label{eq:def_MSE_est}
%\end{align}
%%
%where $\var{ \widehat{U} } = \mean{(\widehat{U} - \mean{\widehat{U}})^2}$, 
%$\mathrm{Bias}(\widehat{U}, U) = \mean{\mean{\widehat{U}} - U}$, 
%$\langle \cdot \rangle$ denotes the ensemble average,
%and $U$ is the value to be estimated.
%For a sample of gamma distributed and independent random variables 
%$\{x_i\}_{i=1}^{N}$, $\mean{\estmu} = \mu = \gamma \theta$
%and $\mean{\estvar} = \mu_2 = \gamma \theta^2$ holds.
%Thus the estimators defined in \Eqnref{est_mu_var} are unbiased and their 
%mean-square error is given by their variance, $\var{ \estmu }$ and 
%$\var{ \estvar }$. 
%The rate of convergence of the estimators given in \Eqnref{est_mu_var}
%and \Eqnref{def_est_SG_KG}, to their asymptotic values as a function of
%sample length is thus given by their mean square error. 
%
%\footnote{In a time series, $\{x_{i}\}$ are usually correlated, we neglect this for
%simplicity.}

With $\gamma = \mu^2 / \sigma^2$, the mean squared error on the estimators for sample mean 
and variance, given in \Eqnref{est_mu_var}, can be propagated on to a mean-square error on
\Eqnref{def_est_SG_KG} using Gaussian propagation of uncertainty:
\begin{align}
    \mse{\estskwG} & = 4 \frac{\estvar}{\estmu^4} \msemu + \frac{1}{\estvar \estmu^2} \msevar - 4 \frac{1}{\estmu^3} \covmusigma, \label{eq:MSE_S} \\
    \mse{\estkrtG} & = 144 \frac{\estvar^2}{\estmu^6} \msemu + 36 \frac{1}{\estmu^4} \msevar - 144 \frac{\estvar}{\estmu^5} \covmusigma. \label{eq:MSE_K}
\end{align}
Here $\mathrm{COV}(\widehat{A}, \widehat{B}) = \mean{(\widehat{A} - \mean{A})(\widehat{B} - \mean{B})}$.
Thus, the mean squared errors on estimators for coefficients of skewness and 
excess kurtosis can be expressed through the mean squared errors on the mean
and variance, and through the covariance between $\estmu$ and $\estvar$. 

We now proceed to derive analytic expressions for $\mse{\estmu}$ and $\mse{\estvar}$.
With the definition of $\widehat{\mu}$ in \Eqnref{est_mu_var}, and using 
$\mean{\est{\mu}} = \mu = \mean{\Phi(t)}$, we find
\begin{align}
    \msemu = \mean{\left(\estmu - \mu \right)^2} = -\mean{\Phi(t)}^2 + \frac{1}{N^2} \sum\limits_{i=1}^{N} \sum\limits_{j=1}^{N} \mean{\Phi(t_i)\Phi(t_j)}.
    \label{eq:MSE_stop1}
\end{align}

\begin{figure}[htb]
\begin{minipage}{0.95\textwidth}
    \includegraphics[width=\textwidth]{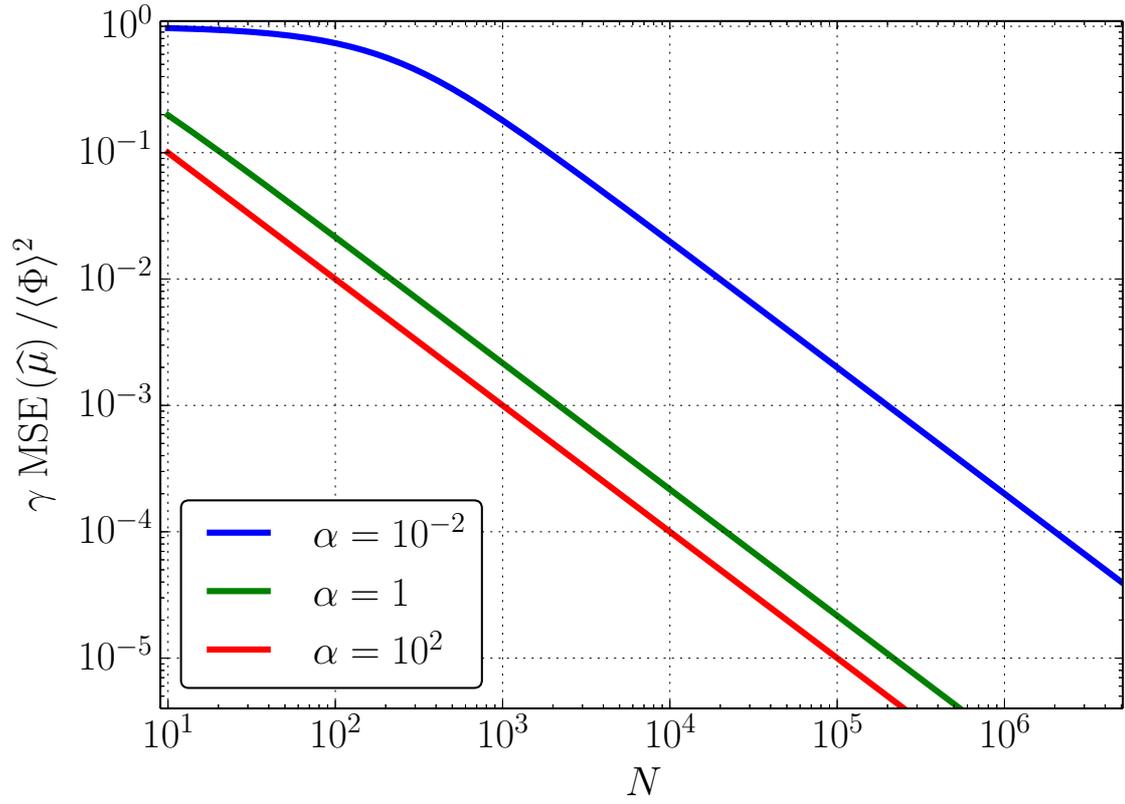}
    \caption{Relative mean squared error on $\estmu$, given by \Eqnref{MSE_mu}, as a function of 
        the number of data points $N$ for three values of the normalized sampling rate $\alpha = \triangle_t / \taud$.}
    \label{fig:mse_mu_alpha}
\end{minipage}
\end{figure}

In order to evaluate the sum over the discrete correlation function, we evaluate
the continuous two-point correlation function given by \Eqnref{sn_corr_twopoint}
at the discrete sampling times, with a discrete time lag given by
$\tau = \tau_{ij} = t_i - t_j$. This gives 
\begin{align*}
    \msemu & = \frac{1}{N} \mean{A}^2 \taudw\left[1 + \frac{1}{N} \sum\limits_{\substack{i,j=1 \\ i \neq j}}^{N} \bigexp{-\frac{|\tau_{ij}|}{\taud}}\right].
\end{align*}
Defining $\alpha = \triangle_t / \taud$, we evaluate the sum as a geometric series, 
\begin{align}
    \frac{1}{2} \sum\limits_{\substack{i,j=1 \\ i \neq j}}^{N} \bigexp{-\frac{|\tau_{ij}|}{\taud}} 
    & =\frac{N + e^{-\alpha N} -1 - Ne^{-\alpha}}{2 \sinh^2 \left(\alpha / 2\right)},
    \label{eq:sumij_result}
\end{align}
to find the mean squared error
\begin{align}
    \msemu & = \frac{1}{N} \mean{A}^2 \frac{\taud}{\tauw} \left[ 1 + \frac{1}{N} \frac{N + e^{-\alpha N} - 1 - N e^{-\alpha}}{2 \sinh^2 \left(\alpha / 2\right) } \right].
    \label{eq:MSE_mu}
\end{align}
\Figref{mse_mu_alpha} shows the normalized mean squared error as a function of the 
of sample size, $N$.
The parameter $\alpha$ relates the sampling time to the pulse duration time.
For $\alpha \gg 1$, the obtained samples are uncorrelated, while the limit $\alpha \ll 1$
describes the case of high sampling frequency where the time series is well resolved
on the time scale of the individual pulses. We find for the corresponding limits
\begin{align}
    \msemu = \frac{1}{N} \mean{\Phi(t)}^2 \frac{\tauw}{\taud} \times
    \begin{cases}
        1 \quad & \alpha \gg 1, \\
        1 + \frac{2}{N} \frac{e^{-\alpha N} - \left(1 - \alpha N \right)}{\alpha^2} \quad & \alpha \ll 1. \label{eq:MSE_mu_lim}
    \end{cases}
\end{align}
For both limits, $\msemu$ is proportional to $\mu^2$ and inversely proportional 
to the intermittency parameter $\gamma = \taud / \tauw$.

In the case of low sampling frequency, $\alpha \gg 1$, the mean squared error on the 
estimator of the mean becomes independent of the sampling frequency and is only 
determined by the parameters of the underlying shot noise process. In this case, the 
relative error $\msemu / \mean{\Phi}^2$ is inversely proportional to $\gamma$ and the 
number of data points $N$. Thus, a highly intermittent process, $\gamma \ll 1$, 
features a larger relative error on the mean than a process with significant pulse overlap,
$\gamma \gg 1$. 
In the case of high sampling frequency, $\alpha \ll 1$, finite correlation
effects contribute to the mean squared error on $\estmu$, given by the
non-canceling terms of the series expansion of $\exp (- \alpha N )$ in \Eqnref{MSE_mu_lim}.
Continuing with the high sampling frequency limit, we now further take the limit
$\alpha N \gg 1$. This describes the case of a total sample time long compared
to the pulse duration time, $T = N \triangle_t \gg \taud$.
In this case the mean square error on the mean is given by
\begin{align}
    \msemu = \frac{2}{\alpha N} \mean{\Phi(t)}^2 \; \frac{\tauw}{\taud}.
    \label{eq:MSE_mu_lim_alphaN}
\end{align}
As in the low sampling frequency limit, the mean square error on $\mu$ converges
as $N^{-1}$, but is larger by a factor of $2 / \alpha$, where $\alpha$ was
assumed to be small.

In \Figref{mse_mu_alpha} we present $\msemu\,$ for $\alpha = 10^{-2}$, $1$, and $10^{2}$.
The first value corresponds to the fast sampling limit, the second value corresponds
to sampling on a time scale comparable to the decay time of an individual pulse and
the third value corresponds to sampling on a slower time scale.
The relative error for the case $\alpha \ll 1$ is clearly largest. For $N \lesssim 10^4$, the $N$ 
dependency of $\msemu$ is weaker than $N^{-1}$. Increasing $N$ to $N \gtrsim 10^{4}$ gives 
$\alpha N \gg 1$, such that $\msemu \sim 1/N$ holds. For $\alpha = 1$, and $\alpha = 10$, 
$\alpha N \gg 1$ holds, and we find that the relative mean squared error on the mean is 
inversely proportional to the number of samples $N$, in accordance with \Eqnref{MSE_mu_lim}. 

We note here, that instead of evaluating the geometrical sum that leads to
\Eqnref{sumij_result} explicitly, it is more
convenient to rewrite the sum over the correlation function in \Eqnref{MSE_stop1} as
a Riemann sum and approximate it as an integral:
\begin{align}
    \sum\limits_{i\neq j} e^{-\alpha|i-j|} \simeq \int\limits_{0}^{N} \mathrm{d}i\; \int\limits_{0}^{N} \mathrm{d}j
        \left[ \Theta(i-j) e^{\alpha(j-i)} + \Theta(j-i) e^{\alpha(i-j)} \right]
        = 2\frac{\alpha N + e^{-\alpha N} - 1}{\alpha^2}.
    \label{eq:intij_result}
\end{align}
For the approximation to be valid, it is required that $\mathrm{d}i / N, \mathrm{d}j / N \ll 1$,
and that the variation of the integrand over $\triangle_i \times \triangle_j$
must be small, $\alpha \ll 1$.
%or alternatively $N \gg 1$. Second, the variation of the integrand over 
%$\triangle_i \times \triangle_j$ must be small which requires $\alpha \ll 1$.
Approximating the sum as in \Eqnref{intij_result} therefore yields the 
same result for $\msemu$ as the limit $\alpha \gg 1$ given in \Eqnref{MSE_mu_lim}.

Expressions for the mean squared error on the estimator $\estvar$ and the covariance
$\covmusigma$ are derived using the same approach as used to derive \Eqnref{MSE_mu}.
With $\mse{\estvar} = \mean{(\estvar - \sigma^2)^2}$, and
$\covmusigma = \mean{(\estmu - \mu) (\estvar - \sigma^2)}$, it follows from \Eqnref{est_mu_var} 
that expressions for summations over third and fourth order correlation functions of
the signal given by \Eqnref{def_shotnoise} have to be evaluated to obtain closed expressions.
Postponing the details of these calculations to the appendix, we present here only 
the resulting expressions. The mean squared error on the variance is given by
\begin{align}
\msevar = \mean{A}^4 
    \left[ \left(\taudw\right)^2 \left( \frac{2}{\alpha N} + \frac{-5 -8 e^{-\alpha N} + e^{-2\alpha N}}{\alpha^2 N^2} \right) \right. \nonumber \\
    \left. \quad + \taudw \left( \frac{6}{\alpha N} + \frac{-27 + 3 e^{-2 \alpha N} }{\alpha^2 N^2} \right) \right] 
    \quad + \mathcal{O} \left( N^{-3} \right),
\label{eq:MSE_var}
\end{align}
while the covariance between the estimators of the mean and variance is given by
\begin{align}
    \covmusigma = \mean{A}^3 \left[ \left( \taudw \right)^2 4 \frac{1 - e^{-\alpha N}}{\alpha^2 N^2} + \taudw \left( \frac{3}{\alpha N} + \frac{-17 + 4 e^{-\alpha N} - 4 e^{-2 \alpha N}} {2 \alpha^2 N^2} \right. \right. \nonumber \\
    \left. \left. + \frac{9 - 12 e^{-\alpha N} + 3 e^{- 2 \alpha N}}{\alpha^3 N^3} \right) \right].
\label{eq:COV_mu_var}
\end{align}
The results, given in \Eqsref{MSE_mu}, \eqref{eq:MSE_var}, and \eqref{eq:COV_mu_var}, are
finally used to evaluate \Eqsref{MSE_S}, and \eqref{eq:MSE_K}, yielding the mean squared error
on $\estskwG$ and $\estkrtG$.
The higher order terms in \Eqnref{MSE_var} are readily calculated by the method
described in appendix \ref{sec:appA} and are not written out here due to space restrictions. 

In the limit $\alpha N \gg 1$, leading order terms in \Eqsref{MSE_var} and 
\eqref{eq:COV_mu_var} are inversely proportional to $\alpha N$:
\begin{align}
    \covmusigma & = \frac{3}{\alpha N} \mean{\Phi(t)} \var{\Phi(t)} \tauwd \label{eq:COV_mu_var_lim_alphaN} \\
    \msevar & = \frac{2}{\alpha N} \var{\Phi(t)}^2 \left( 1 + 3 \frac{\tauw}{\taud} \right) \label{eq:MSE_var_lim_alphaN}.
\end{align}
While \Eqsref{MSE_mu_lim_alphaN} and \eqref{eq:COV_mu_var_lim_alphaN} are proportional 
to $\gamma$, $\msevar$ depends also quadratically on $\gamma$.

\section{Comparison to synthetic time series}
\label{sec:synth_ts}
In this section we compare the derived expressions for the mean squared error on the
estimators for the sample mean, variance, skewness, and kurtosis, against sample variances 
from the respective estimators computed of synthetic time series of the stochastic
process given by \Eqnref{def_shotnoise}. 

\begin{figure}[htb]
\begin{minipage}{0.95\textwidth}
    \includegraphics[width=\textwidth]{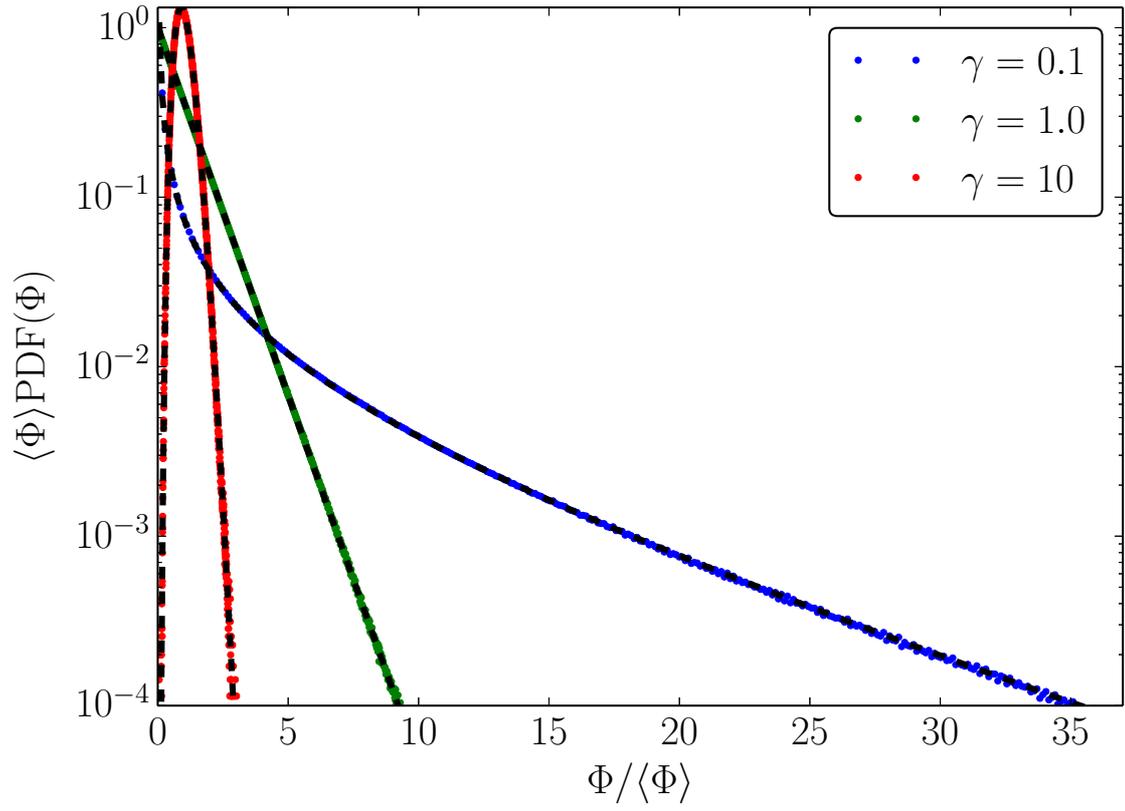}
    \caption{Histogram of synthetic time series with $\gamma = 0.1$, $1.0$, and $10$. Overlaid
        (black dashed lines) is the Gamma distribution given by \Eqnref{gamma_PDF} with a scale parameter $\theta = 1$.}
    \label{fig:sn_hist}
\end{minipage}
\end{figure}

To generate synthetic time series, the number of pulses $K$, the pulse duration
time $\taud$, the intermittency parameter $\gamma$, the pulse amplitude scale $\mean{A}$,
and sampling time $\triangle_t$ are specified. The total number of samples in the time 
series is given by $N = K / \gamma \triangle_t$.
The pulse arrival times $t_k$ and pulse amplitudes $A_k$, $k = 1 \ldots K$, are drawn 
from a uniform distribution on $[0: K/\gamma]$ and from 
$P_A(A) = \exp \left(-A / \mean{A}\right) / \mean{A}$ respectively. The tuples $(t_k, A_k)$
are subsequently sorted by arrival time and the time series is generated according
to \Eqnref{def_shotnoise} using the exponential pulse shape given by \Eqnref{pulse_shape_exp}. 
The computation of the time series elements is implemented by a parallel algorithm utilizing
graphical processing units. 
For our analysis we generate time series for $\gamma = 0.1$ and $10$,
$\triangle_t = 0.01$, and time and amplitude normalized such that $\taud = 1$
and $\mean{A} = 1$.
Thus, $\alpha = \triangle_t / \taud = 0.01$ for both time series. Both time series
have $N = 10^8$ samples, which requires $K = 10^5$ for the time series with $\gamma = 0.1$
and $K = 10^7$ for the time series with $\gamma = 10$. The histogram for both time 
series is shown in \Figref{sn_hist}.

Each time series generated this way is a realization of the stochastic process 
described by \Eqnref{def_shotnoise}. We wish to estimate the lowest order statistical 
moments, as well as their mean squared errors, of these time series as a function of the
sample size. For this, we partition the time series for a given value of $\gamma$ into 
$\mathcal{M}$ equally long sub-time series with $N_\mathcal{M} = N / \mathcal{M}$ elements each. 
The partitioned sample size $N_\mathcal{M}$ is varied from $2 \times 10^3$ to $10^6$ elements
as to partition the total time series into $\mathcal{M} \in \{100, 200, 500, \ldots, 50000\}$ 
sub-time series.

For each sub-time series, we evaluate the estimators \Eqnref{est_mu_var} and \Eqnref{def_est_SG_KG},
which yields the sets 
$\{\estmu_m\}$,
$\{\estvar_m\}$,
$\{\estskwGm\}$, and
$\{\estkrtGm\}$, with $m \in (1, \ldots \mathcal{M})$.
The variance of these sets of estimators is then compared to the analytic expressions
for their variance, given by \Eqsref{MSE_mu}, \eqref{eq:MSE_var}, \eqref{eq:MSE_S}, and \eqref{eq:MSE_K}. 
%With the chosen partition of $N$ in $\mathcal{S}$ sub time-series, the sets $\{ \est{A}_s \}_{s=1}^{\mathcal{S}}$
%have the same number of elements for any given $N_\mathcal{S}$.
%
Additionally, we wish to compare the precision and accuracy of the proposed estimators 
given by \Eqnref{def_est_SG_KG} to the estimators defined by the method of moments 
in \Eqnref{est_SKmom}. For this, we also evaluate \Eqnref{est_SKmom} on each sub time-series 
and compute the sample average  and variance of the resulting set of estimators. 

\begin{figure}[htb]
\begin{minipage}{0.95\textwidth}
    \includegraphics[width=\textwidth]{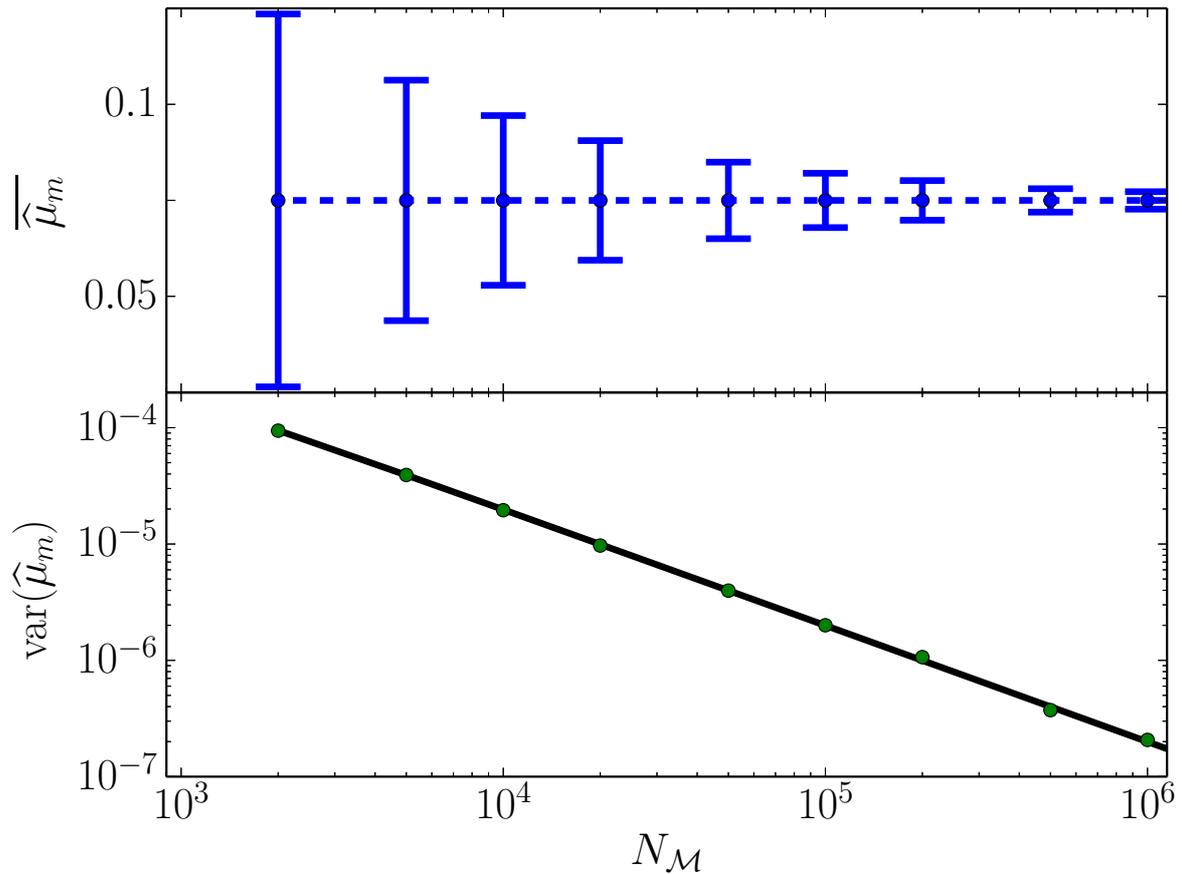}
    \caption{Sample mean (upper panel) and variance (lower panel) of the estimators 
        $\{\estmu_m\}$ as a function of the partitioned sample size $N_\mathcal{M}$, computed from the synthetic 
        time series with $\gamma = 0.1$. The dashed line in the upper panel is $\estmu$ computed 
        with $N$ data points, the black line in the lower panel is given by \Eqnref{MSE_mu}.}
    \label{fig:gamma01_mean}
\end{minipage}
\end{figure}

\begin{figure}[htb]
\begin{minipage}{0.95\textwidth}
    \includegraphics[width=\textwidth]{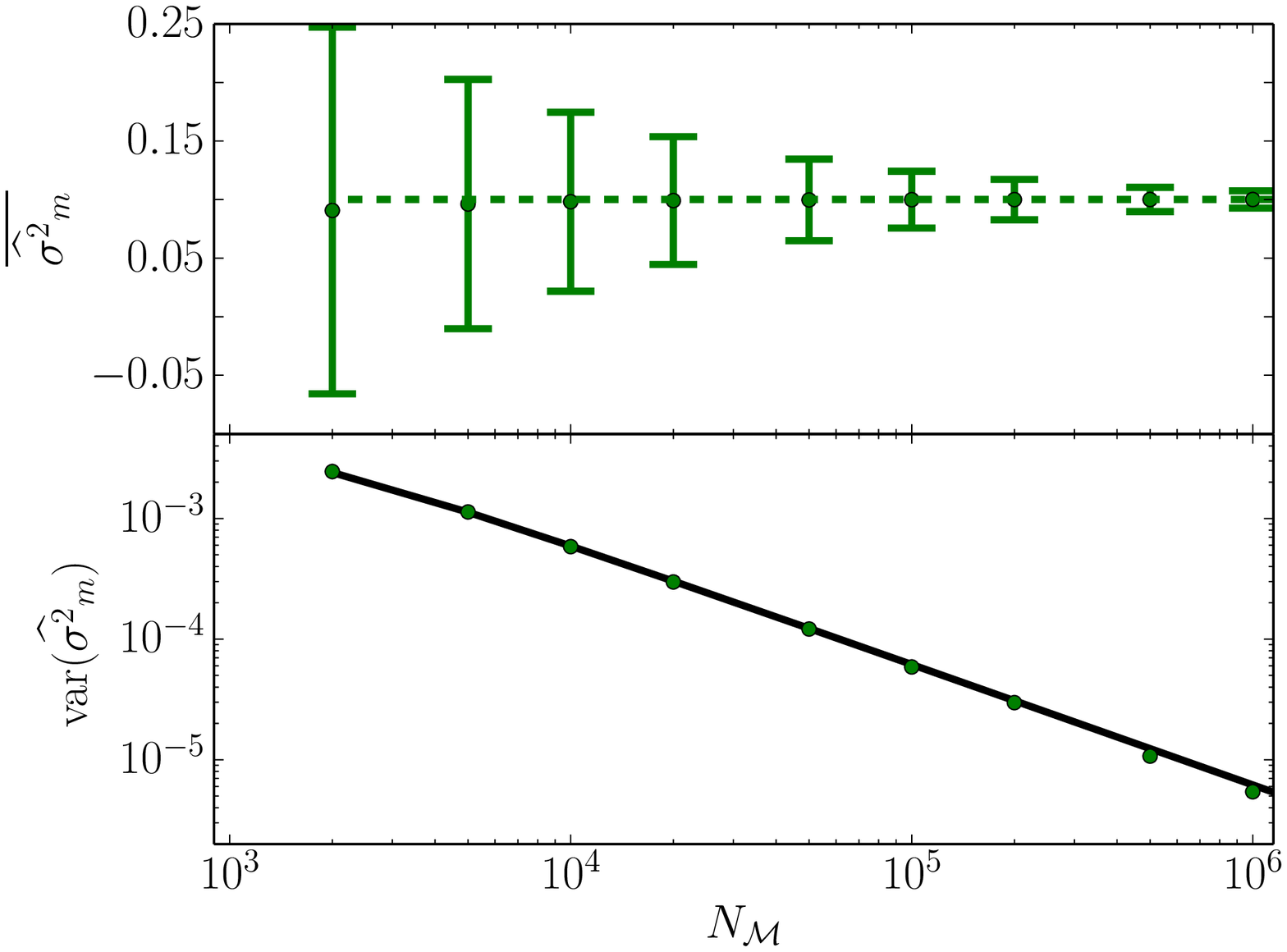}
    \caption{Sample mean (upper panel) and variance (lower panel) of the estimators 
        $\{\estvar_m\}$ computed from the synthetic time series with $\gamma = 0.1$.
        The dashed line in the upper panel is $\estvar$ computed with $N$ data points, 
        the black line in the lower panel is given by \Eqnref{MSE_var}.}
    \label{fig:gamma01_var}
\end{minipage}
\end{figure}

\begin{figure}[htb]
\begin{minipage}{0.95\textwidth}
    \includegraphics[width=\textwidth]{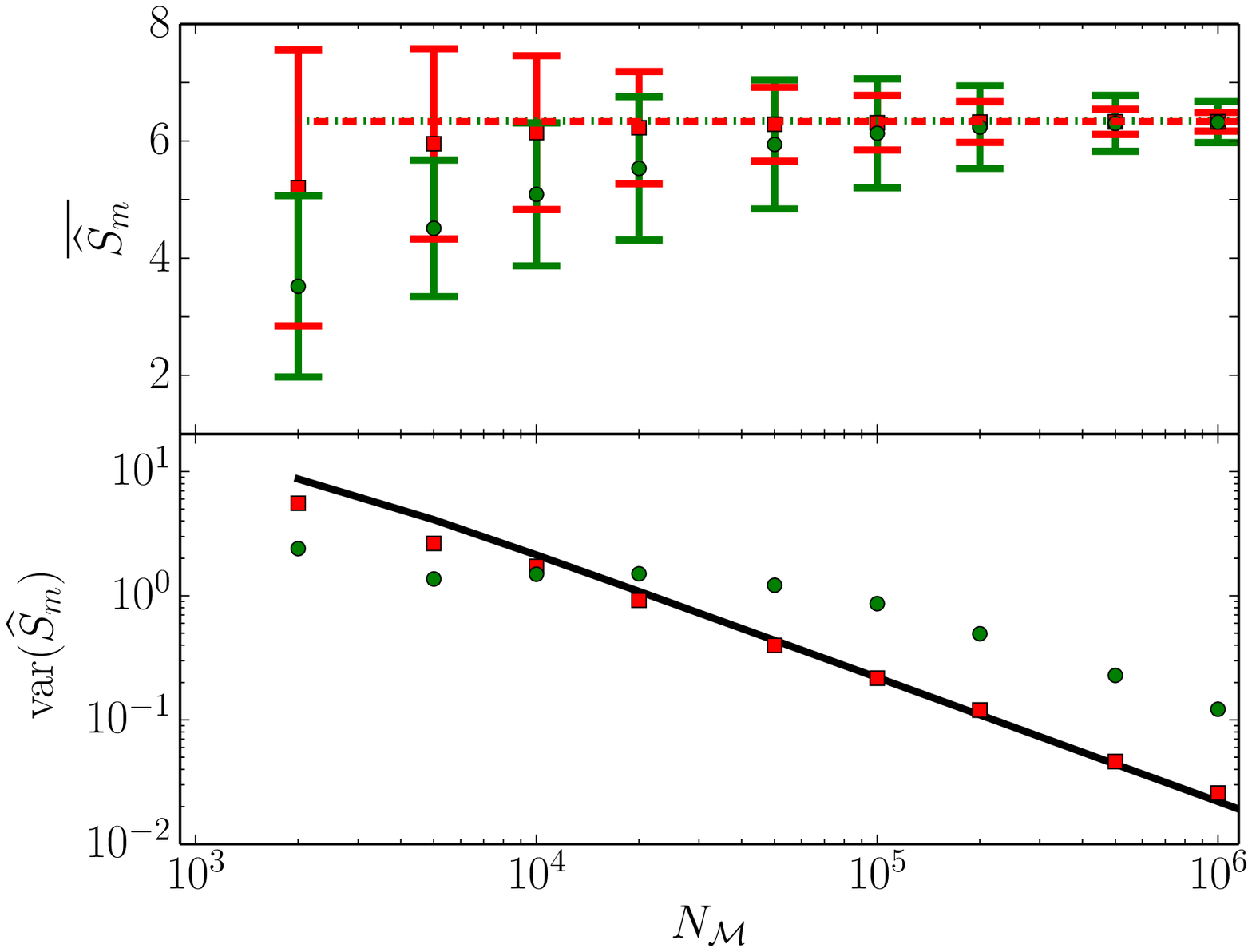}
    \caption{Sample mean (upper panel) and variance (lower panel) of the estimators
        $\{\estskwGm\}$ (red square) and $\{\estskwm\}$ (green circle) computed from 
        the synthetic time series with $\gamma = 0.1$.
        The dashed (dotted) line in the upper panel is $\estskwG$ ($\estskw$) 
        computed with $N$ data points, the black line in the lower panel is given 
        by \Eqnref{MSE_S}.}
    \label{fig:gamma01_skew}
\end{minipage}
\end{figure}

\begin{figure}[htb]
\begin{minipage}{0.95\textwidth}
    \includegraphics[width=\textwidth]{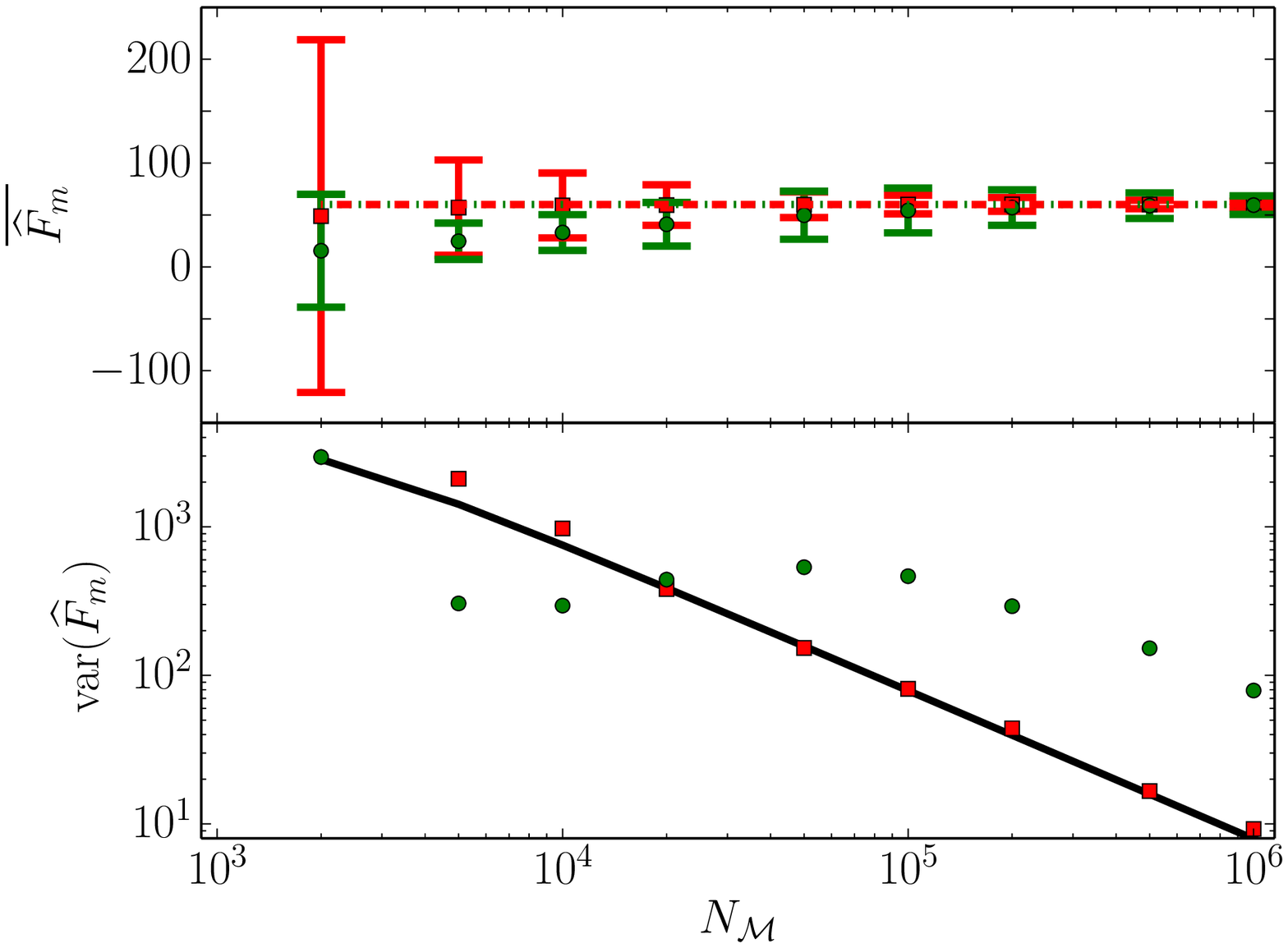}
    \caption{Sample mean (upper panel) and variance (lower panel) of the estimators
        $\{\estkrtGm\}$ (red square) and $\{\estkrtm\}$ (green circle) computed from 
        the synthetic time series with $\gamma = 0.1$.
        The dashed (dotted) line in the upper panel is $\estkrtG$ ($\estkrt$) 
        computed with $N$ data points, the black line in the lower panel is given 
        by \Eqnref{MSE_K}.}
    \label{fig:gamma01_kurt}
\end{minipage}
\end{figure}

\Figsref{gamma01_mean} - \ref{fig:gamma01_kurt} show the results of this comparison for
the synthetic time series with $\gamma=0.1$.
%
%%%%%%%%%%%%%%%%%%%%%%%%%%%%%%%%%%%%%%%%%%%%%%%%%%%%%%%%%%%%%%%%%%%%%%%%%%%%%%%%%%%%%%%%%%
% gamma = 0.1, mean
The upper panel in \Figref{gamma01_mean} shows the sample average of $\{ \estmu_m \}$ 
with error bars given by the root-mean square of the set for a given sample size 
$N_\mathcal{M}$. Because $\estmu$ is linear in all its arguments $x_i$ the sample average 
of $\{\estmu_m\}$ for any given $N_\mathcal{M}$ equals $\estmu$ computed for the entire 
time series.
The lower panel compares the sample variance of $\{ \estmu_m \}$ for a given $N_\mathcal{M}$
to that given by \Eqnref{MSE_mu}. For the presented data, the long sample limit applies since
$\alpha N_\mathcal{M} \ge 20 \gg 1$. 
A least squares fit on $\textvar{\{ \estmu_m \}}$ shows a dependence of $\sim N_\mathcal{M}^{-0.90}$ 
which agrees with the analytical result of $\msemu \sim N_\mathcal{M}^{-1}$, given by
\Eqnref{MSE_mu_lim_alphaN}. 

%%%%%%%%%%%%%%%%%%%%%%%%%%%%%%%%%%%%%%%%%%%%%%%%%%%%%%%%%%%%%%%%%%%%%%%%%%%%%%%%%%%%%%%%%%
% gamma = 0.1, var
In \Figref{gamma01_var} we present the sample average of the estimators $\{ \estvar_m \}$ 
with error bars given by the root-mean square of the set of estimators for a given sample 
size $N_\mathcal{M}$. We find that the sample variance of the estimators compare well with 
the analytic result given by \Eqnref{MSE_var}. A least squares fit reveals that 
$\textvar{\{ \estvar_m \} } \sim N_\mathcal{M}^{-0.91}$ while \Eqnref{MSE_var} behaves 
as $N_\mathcal{M}^{-1}$.
%
%%%%%%%%%%%%%%%%%%%%%%%%%%%%%%%%%%%%%%%%%%%%%%%%%%%%%%%%%%%%%%%%%%%%%%%%%%%%%%%%%%%%%%%%%%
% gamma = 0.1, skew
The sample averages of the skewness estimators $\{ \estskwGm \}$, \Eqnref{def_est_SG_KG}, 
and $\{ \estskwm \}$, \Eqnref{est_SKmom}, as a function of sample size are shown 
in the upper panel of \Figref{gamma01_skew}. 
Both estimators yield the same coefficient of skewness when applied to the entire 
time series and converge to this coefficient with increasing $N_\mathcal{M}$. 
For a small number of samples, $N_\mathcal{M} \lesssim 10^4$, the estimator based on the method 
of moments estimates a sample skewness that is on average more than one standard
deviation from the true value of skewness.
Again, the error bars are given by the root mean square value of the set of estimators for 
any $N_\mathcal{M}$.
For larger samples $\textvar{ \{ \estskwGm \} }$ is smaller than $\textvar{ \{ \estskwm \} }$ 
by about one order of magnitude and both are inversely proportional to the number of samples.
\Eqnref{MSE_S} yields $\mse{\estskwG} \sim N_\mathcal{M}^{-0.99}$ which 
compares favorably to the dependency of the sample variance of the estimator
based on the method of moments on the number of samples, 
$\textvar{\{ \estskwGm \}} \sim N_\mathcal{M}^{-1.00}$.
%
%%%%%%%%%%%%%%%%%%%%%%%%%%%%%%%%%%%%%%%%%%%%%%%%%%%%%%%%%%%%%%%%%%%%%%%%%%%%%%%%%%%%%%%%%%
% gamma = 0.1, kurt
The discussion of the skewness estimators applies similarly to the kurtosis estimators. 
Intermittent bursts in the time series with $\gamma = 0.1$ cause large deviations from the
time series mean which results
in a large coefficient of excess kurtosis. Dividing the total time series in sub time series
results in large variation of the sample excess kurtosis. 
For samples with
$N_\mathcal{M} \lesssim 10^{4}$ the estimator based on the method of moments performs better than the 
estimator defined in \Eqnref{def_est_SG_KG}. The opposite is true for samples with
$N_\mathcal{M} \gtrsim 10^{4}$, where $\estkrtG$ performs significantly better than $\estkrt$.
In the latter case, $\textvar{ \{\estkrtGm\} }$ is lower than $\textvar{\{ \estkrtm \} }$ by
one order of magnitude.
Both estimators, $\estkrt$ and $\estkrtG$, converge to their full sample estimate which is identical. 
A least squares fit reveals that $\textvar{ \{ \estkrtGm \}} \sim  N_\mathcal{M}^{-1.00}$
while a least-squares fit on \Eqnref{MSE_K} finds a dependency of the form $\sim N_\mathcal{M}^{-0.97}$. 

\begin{figure}
\begin{minipage}{0.95\textwidth}
    \includegraphics[width=\textwidth]{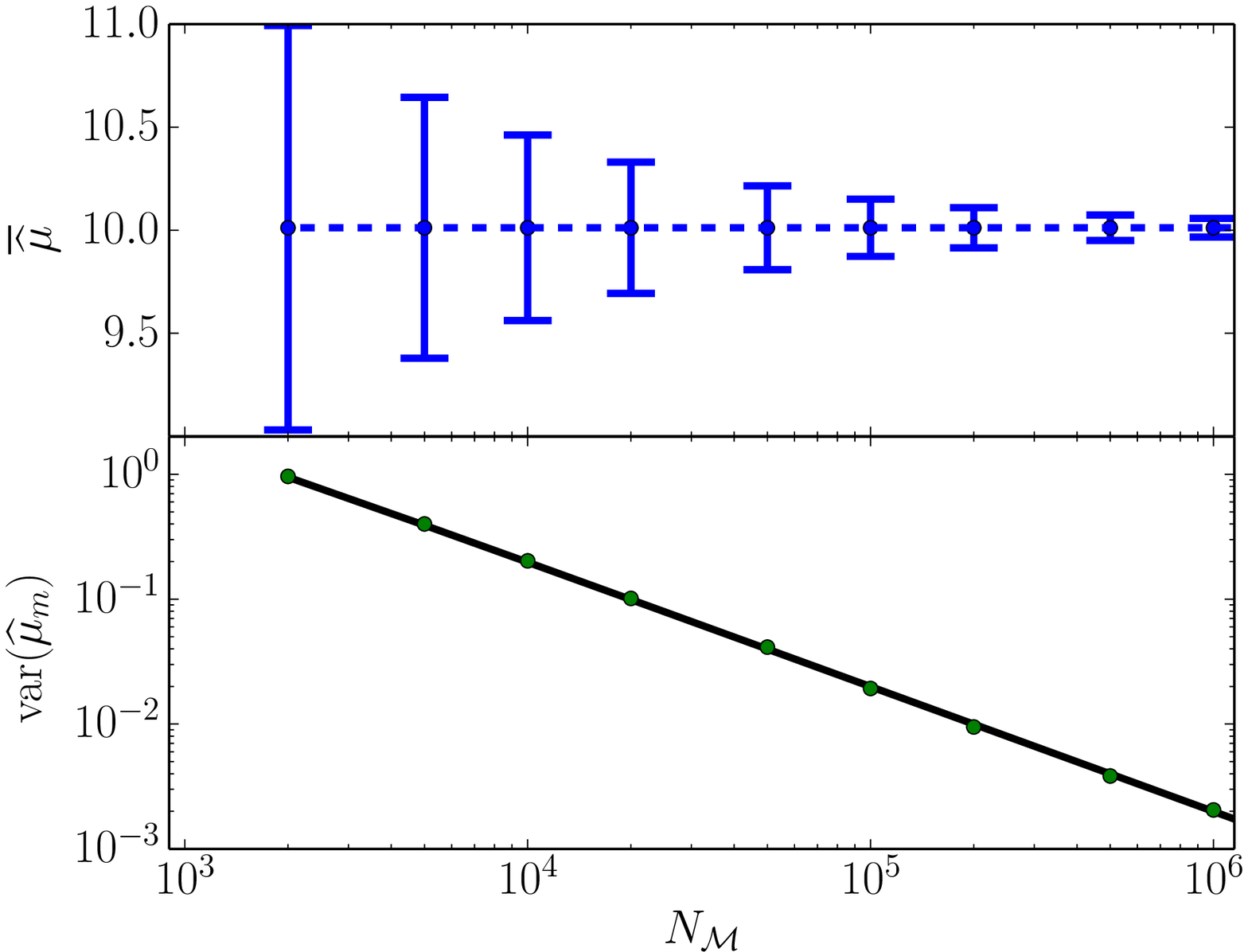}
    \caption{Sample mean (upper panel) and variance (lower panel) of the estimators $\{\estmu_m\}$
        computed from the synthetic time series with $\gamma = 10$.
        The dashed line in the upper panel is $\estmu$ computed with $N$ data points, the 
        black line in the lower panel is given by \Eqnref{MSE_mu}.}
    \label{fig:gamma10_mean}
\end{minipage}
\end{figure}

\begin{figure}[htb]
\begin{minipage}{0.95\textwidth}
    \includegraphics[width=\textwidth]{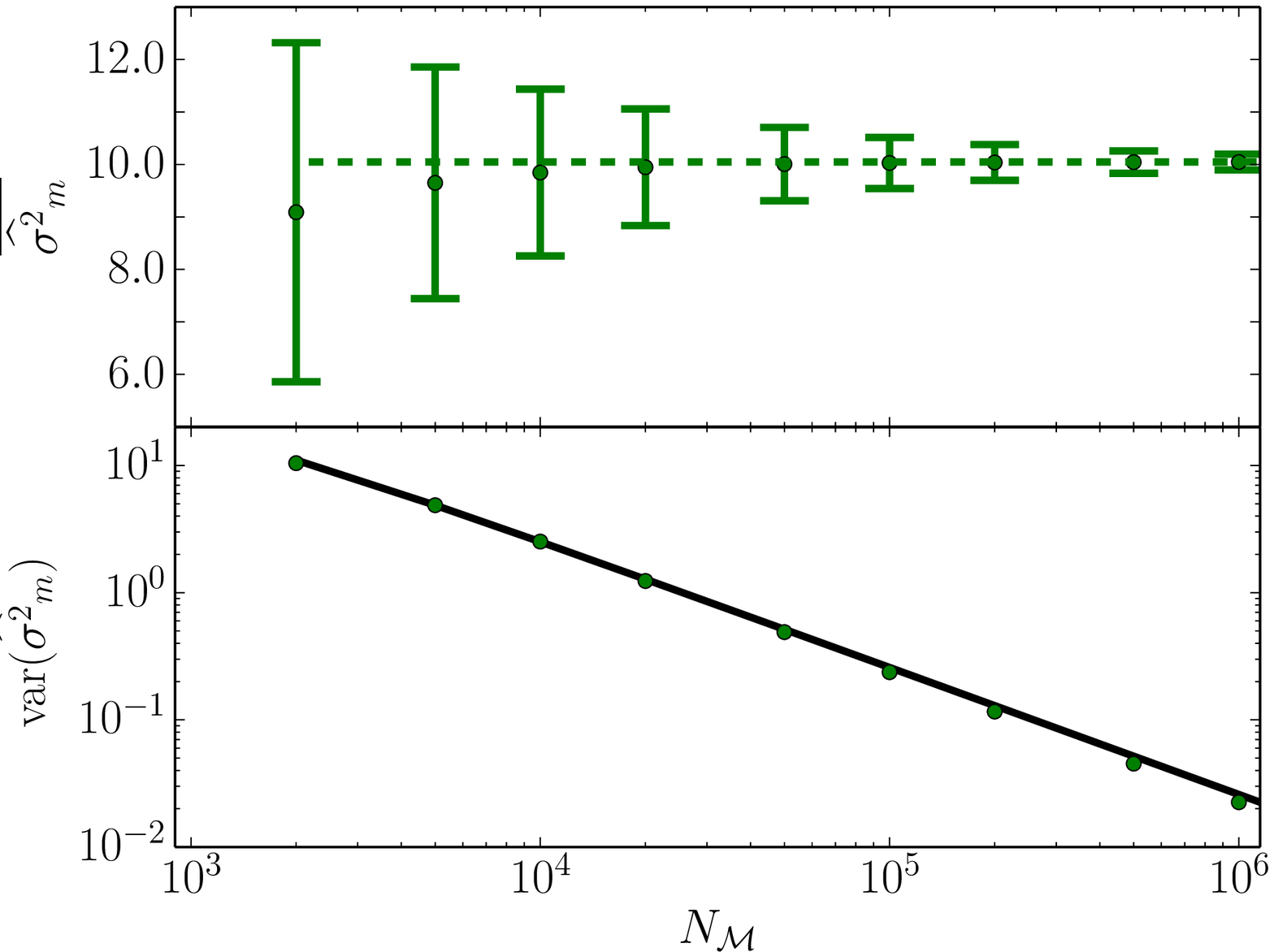}
    \caption{Sample mean (upper panel) and variance (lower panel) of the estimators $\{\estvar_m\}$
        computed from the synthetic time series with $\gamma = 10$.
        The dashed line in the upper panel is $\estvar$ computed with $N$ data points, the 
        black line in the lower panel is given by \Eqnref{MSE_var}.}
    \label{fig:gamma10_var}
\end{minipage}
\end{figure}

\begin{figure}[htb]
\begin{minipage}{0.95\textwidth}
    \includegraphics[width=\textwidth]{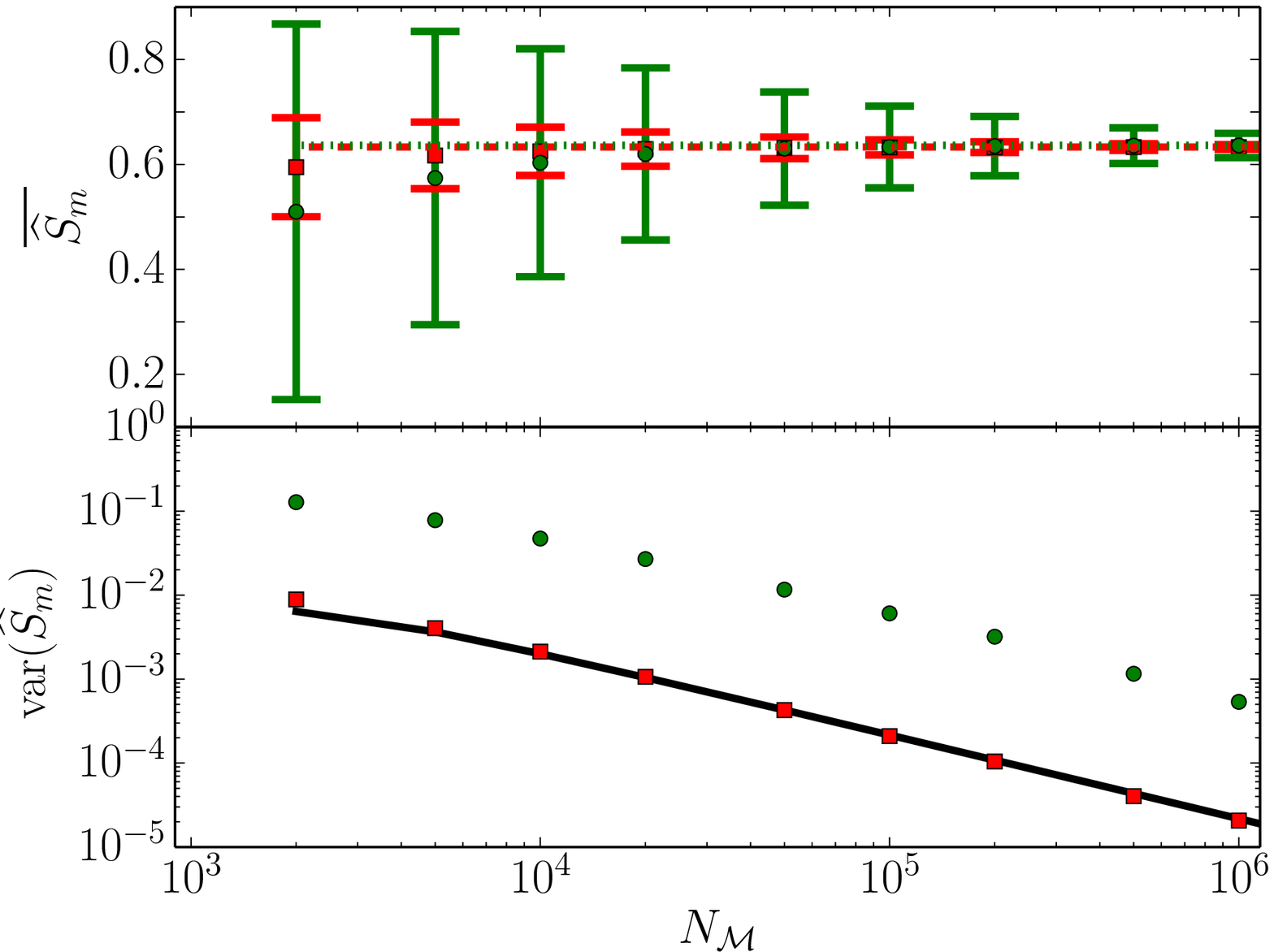}
    \caption{Sample mean (upper panel) and variance (lower panel) of the estimators
        $\{\estskwGm\}$ (red square) and $\{\estskwm\}$ (green circle) computed from the synthetic
        time series with $\gamma = 10$.
        The dashed (dotted) line in the upper panel is $\estskwG$ ($\estskw$) computed with 
    $N$ data points, the black line in the lower panel is given by \Eqnref{MSE_S}.}
    \label{fig:gamma10_skew}
\end{minipage}
\end{figure}

\begin{figure}
\begin{minipage}{0.95\textwidth}
    \includegraphics[width=\textwidth]{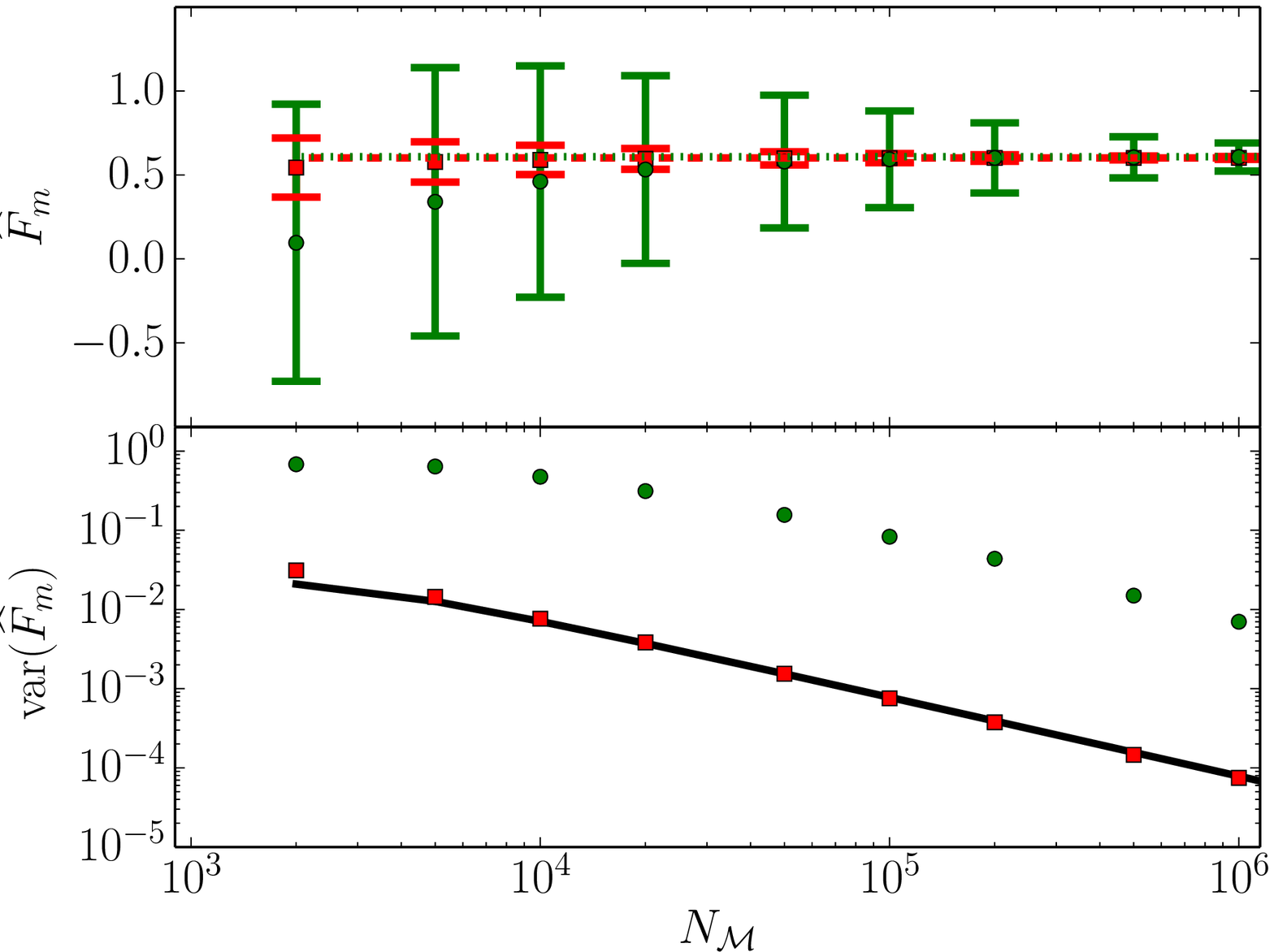}
    \caption{Sample mean (upper panel) and variance (lower panel) of the estimators
        $\{\estkrtGm\}$ (red square) and $\{\estkrtm\}$ (green circle) computed from the synthetic
        time series with $\gamma = 10$.
        The dashed (dotted) line in the upper panel is $\estkrtG$ ($\estkrt$) computed with 
        $N$ data points, the black line in the lower panel is given by \Eqnref{MSE_K}.}
    \label{fig:gamma10_kurt}
\end{minipage}
\end{figure}

%%%%%%%%%%%%%%%%%%%%%%%%%%%%%%%%%%%%%%%%%%%%%%%%%%%%%%%%%%%%%%%%%%%%%%%%%%%%%%%%%%%%%%%%%%
% gamma = 10.0
In \Figsref{gamma10_mean} to \ref{fig:gamma10_kurt} we present the same data analysis 
as in the previous figures, for the time series with a large intermittency parameters, 
$\gamma = 10$. This time series features a large pulse overlap.
Again, with $N_\mathcal{M} \ge 2 \times 10^3$, the limit $\alpha N_\mathcal{M} \gg 1$ applies. 
The lower panel in \Figref{gamma10_mean} shows a good agreement between
\Eqnref{MSE_var} and the empirical scaling of $\{ \estmu_m \}$ which is found by a least
squares fit to be $\textvar{ \{ \estmu_m \}} \sim N_\mathcal{M}^{-0.98}$, 
in good agreement with \Eqnref{MSE_mu_lim_alphaN}. 
We further find that $\textvar{ \{ \estvar_m \} }$ is also inversely proportional to the number of 
samples, see \Figref{gamma10_var}. 
For \Figsref{gamma10_skew} and \ref{fig:gamma10_kurt} we note that the coefficients of skewness 
and  excess kurtosis are one order of magnitude lower for $\gamma = 10$ than for $\gamma = 0.1$,
in accordance with \Eqnref{sn_mom}. 
Due to significant pulse overlap, sample variances of skewness and excess kurtosis 
show a smaller variance than in the case of $\gamma = 0.1$. Again, the magnitude of 
$\textvar{ \{ \estskwm \} }$, and $\textvar{ \{ \estkrtm \}}$ is one order of magnitude
larger than $\textvar{ \{ \estskwGm \} }$, and $\textvar{ \{ \estkrtGm \} }$, respectively,
and the variance of all estimators is approximately inversely proportional to $N_\mathcal{M}$.
For sample sizes up to $N_\mathcal{M} \simeq 10^{4}$, 
$\estkrt$ yields negative values for the sample excess kurtosis, while the 
of excess kurtosis as calculated from the entire sample is positive. This is due to the 
large sample variance of this estimator and a coefficient of excess kurtosis of the 
underlying time series.

%%%%%%%%%%%%%%%%%%%%%%%%%%%%%%%%%%%%%%%%%%%%%%%%%%%%%%%%%%%%%%%%%%%%%%%%%%%%%%%%%%%%%%%%%%%%%%%%%%%%%%%%%%%%%%%
%
%%%%%%%%%%%%%%%%%%%%%%%%%%%%%%%%%%%%%%%%%%%%%%%%%%%%%%%%%%%%%%%%%%%%%%%%%%%%%%%%%%%%%%%%%%%%%%%%%%%%%%%%%%%%%%%
%
\section{Discussions and Conclusion}
\label{sec:conclusion}
We have utilized a stochastic model for intermittent particle density fluctuations in 
scrape-off layer plasmas, given in Ref.~\Ref{garcia-2012}, to calculate expressions for the 
mean squared error on estimators of sample mean, variance, coefficients 
of skewness, and excess kurtosis as a function of sample length, sampling 
frequency, and parameters of the stochastic process.
We find that the mean squared error on the estimator of the sample mean is proportional 
to the square of the ensemble average of the underlying stochastic process, inversely 
proportional to the intermittency parameter $\gamma$, and inversely proportional to the 
number of samples, $N$. In the limit of high sampling frequency and large number of
samples, the mean squared error also depends on the ratio of the pulse decay time to
sampling frequency, as given by \Eqnref{MSE_mu_lim_alphaN}. 

The derived expressions for the mean squared error on the estimator for the sample variance
and covariance between $\estmu$ and $\estvar$ are polynomials in both $\gamma$ and
$N$. These expressions further allow to compute the mean squared error on
the sample skewness and excess kurtosis by inserting them into \Eqsref{MSE_S} and \eqref{eq:MSE_K}.
In the limit of high sampling frequency and large number of samples, we find that
the expressions for $\msemu$ and $\covmusigma$ to be inversely proportional to both,
$N$, and $\alpha$, and to depend on the intermittency parameter $\gamma$.  

We have generated synthetic time series to compare the sample variance of
the estimators for sample mean, variance, skewness and excess kurtosis to the
expressions for their mean squared error. For a large enough number samples,
$\alpha N \gg 1$, all estimators are inversely proportional to $N$.
We further find that estimators for skewness and excess kurtosis,
as defined by \Eqnref{def_est_SG_KG}, allow a more precise and a more accurate
estimation of the sample skewness and kurtosis than estimators based 
on the method of moments given by \Eqnref{est_SKmom}.

The expressions given by \Eqsref{MSE_mu}, \eqref{eq:MSE_var}, \eqref{eq:MSE_S}, and \eqref{eq:MSE_K}
may be directly applied to assess the relative error on sample coefficients 
of mean, variance, skewness, and kurtosis for time series of particle density
fluctuations in tokamak scrape-off layer plasmas. 
We exemplify their usage for a particle density time series that is sampled with
$1 / \triangle_t = 5\, \mathrm{MHz}$ for $T = 2.5\, \mathrm{ms}$ as to obtain
$N = 12500$ samples. Common fluctuation levels in the scrape-off layer are given
by $\Phi_\mathrm{rms} / \mean{\Phi} \approx 0.5$. Using \Eqnref{sn_mom_mean_sigma}
and $\gamma = \taud / \tauw$, this gives $\gamma \approx 4$. Conditional averaging of
the the bursts occurring in particle density time series reveals an exponentially
decaying burst shape with a typical e-folding time of approximately $20\, \mu\mathrm{s}$, so that
$\alpha \approx 0.01$. Thus, the individual bursts are well resolved on the time scale on which the 
particle density is sampled and the assumption $\alpha N \gg 1$ is justified. 
From \Eqnref{MSE_mu_lim_alphaN}, we then compute the relative mean squared error on 
the sample average to be
$\msemu / \mean{\Phi}^2 \simeq 3.2 \times 10^{-3}$
and likewise the relative mean squared error on the sample variance from
\Eqnref{MSE_var_lim_alphaN} to be 
$\msevar / \var{\Phi}^2 \simeq 2.6 \times 10^{-2}$. This translates into relative errors
of approximately $6\%$ on the sample mean and approximately $16\%$ on the sample variance.
The relative mean squared error on skewness and excess kurtosis evaluates to
$\mse{\estskwG} / \estskwG^2 \simeq 8.6\times 10^{-3}$ and
$\mse{\estkrtG} / \estkrtG^2 \simeq 3.8\times 10^{-2}$, which translates into an
relative error of approximately $9\%$ on the sample skewness and approximately $19\%$ on the
sample excess kurtosis. The magnitude of these values is consistent with reported
radial profiles os sample skewness and kurtosis, where the kurtosis profiles show
significantly larger variance than the skewness profiles 
\Ref{garcia-2007-tcv, graves-2005, horacek-2005, garcia-2006-tcv, garcia-2007-coll}.

The expressions for the mean squared error on sample mean, variance, skewness and
kurtosis presented here may be appropriate for errorbars on experimental measurements of particle
density fluctuations, as well as for turbulence simulations of the boundary region
of magnetically confined plasmas.

%The magnitude of these values is consistent with Ref.~\Ref{garcia-2007-tcv}, figures
%(7), and (8), which presents radial profiles of sample skewness and kurtosis, where
%the kurtosis profiles show significantly larger variance than the skewness profiles.

%
\appendix
\section{Derivation of $\msevar$ and $\covmusigma$}
\label{sec:appA}
We start by reminding of the definitions 
$\mathrm{COV}(\est{A}, \est{B}) = \mean{(\est{A} - \mean{A})(\est{B} - \mean{B})}$
and
$\mathrm{var}(\est{B}) = \mean{(\est{B} - \mean{B})^2}$. For $\est{A} = \estmu$ and 
$\est{B} = \estvar$, we evaluate these expressions to be
\begin{align}
    \covmusigma
    & = \frac{1}{N-1} \left( \sum\limits_{i,j=1}^{N} \mean{\Phi(t_i)^2 \Phi(t_j)} - \frac{1}{N^2} \sum\limits_{i,j,k=1}^{N} \mean{\Phi(t_i) \Phi(t_j) \Phi(t_k)} \right) \nonumber \\
    & \quad - \mean{A} \frac{\taud}{\tauw} \frac{1}{N - 1} \left( \sum\limits_{i=1}^{N} \mean{\Phi(t_i)} - \frac{1}{N} \sum\limits_{i,j=1}^{N} \mean{\Phi(t_i) \Phi(t_j) } \right),
    \label{eq:cov_sigmamu_start}
\end{align}
and
\begin{align}
    \mathrm{var}(\estvar)
    & = - \mean{A}^4 \left( \frac{\taud}{\tauw} \right)^2 
    + 4 \mean{A}^4 \left(\frac{\taud}{\tauw}\right)^2 \left( \frac{1}{N^2} \frac{e^{-\alpha N} - (1 - \alpha N)}{\alpha^2} \right) \nonumber \\
    & + \frac{1}{N^2} \left( \sum\limits_{i,j=1}^{N} \mean{\Phi(t_i)^2 \Phi(t_j)^2} \right. 
    \left.  -\frac{2}{N} \sum\limits_{i,j,k=1}^{N} \mean{\Phi(t_i)^2 \Phi(t_j) \Phi(t_k)} \right. \nonumber  \\
& \left. \hspace{10ex} +\frac{1}{N^2} \sum\limits_{i,j,k,l=1}^{N} \mean{\Phi(t_i) \Phi(t_j) \Phi(t_k) \Phi(t_l)  }\right)
    %& + \frac{1}{N^2} \left( \sum\limits_{i,j=1}^{N} \mean{\Phi(t_i)^2 \Phi(t_j)^2} \right. 
    %\left.  -\frac{1}{N} \sum\limits_{i,j,k=1}^{N} \left(\mean{\Phi(t_i)^2 \Phi(t_j) \Phi(t_k)} + \mean{\Phi(t_j) \Phi(t_k) \Phi(t_i)^2   } \right)   \right. \nonumber  \\
%& \left. \hspace{10ex} +\frac{1}{N^2} \sum\limits_{i,j,k,l=1}^{N} \mean{\Phi(t_i) \Phi(t_j) \Phi(t_k) \Phi(t_l)  }\right)
    \label{eq:var_sigma2_start}
\end{align}

We made use of \Eqnref{intij_result} in deriving the last expression. Therefore it is 
only valid in the limit $\alpha \ll 1$. To derive closed expressions for \Eqsref{MSE_S} 
and \eqref{eq:MSE_K} we proceed by deriving expressions for the third- and fourth-order 
correlation functions of the shot noise process \Eqnref{def_shotnoise}.

We start by inserting \Eqnref{def_shotnoise} into the definition of a 
three-point correlation function
\begin{align}
    & \mean{\Phi_K(t) \Phi_K(t + \tau) \Phi_K(t + \tau')} \nonumber \\
    & = \int\limits_{0}^{T} \mathrm{d}t_1 P_t(t_1) \int\limits_{0}^{\infty} \mathrm{d}A_1 P_A(A_1)
            \cdots 
           \int\limits_{0}^{T} \mathrm{d}t_1 PKt(t_K) \int\limits_{0}^{\infty} \mathrm{d}A_K P_A(A_K) \times \nonumber \\
    & \qquad \sum \limits_{p=1}^{K} \sum \limits_{q=1}^{K} \sum \limits_{r=1}^{K} A_p \psi(t - t_p) A_q \psi(t + \tau - t_q) A_r \psi(t + \tau' - t_r) \nonumber \\
    & = \mean{A^3} \sum \limits_{p=q=r=1}^{K} \int \limits_{0}^{T} \frac{\mathrm{d}t_p}{T} \psi(t - t_p) \psi(t + \tau - t_p) \psi(t + \tau' - t_p) \nonumber \\
    % p = q, r
    & \qquad + \mean{A^2}\mean{A} \sum \limits_{p = q = 1}^{K} \sum \limits_{\substack{r = 1 \\ r \neq p}}^{K}  \int \limits_{0}^{T} \frac{\mathrm{d}t_p}{T}  \int \limits_{0}^{T} \frac{\mathrm{d}t_r}{T} 
        \psi(t - t_p) \psi(t + \tau - t_p) \psi(t + \tau' - t_r) \nonumber \\
    % p = r, q
    & \qquad + \mean{A^2}\mean{A} \sum \limits_{p = r = 1}^{K} \sum \limits_{\substack{q = 1 \\ q \neq p}}^{K}  \int \limits_{0}^{T} \frac{\mathrm{d}t_p}{T}  \int \limits_{0}^{T} \frac{\mathrm{d}t_q}{T} 
        \psi(t - t_p) \psi(t + \tau - t_q) \psi(t + \tau' - t_p) \nonumber \\
    % p, q = r
    & \qquad + \mean{A^2}\mean{A} \sum \limits_{q = r = 1}^{K} \sum \limits_{\substack{p = 1 \\ p \neq r}}^{K}  \int \limits_{0}^{T} \frac{\mathrm{d}t_q}{T}  \int \limits_{0}^{T} \frac{\mathrm{d}t_p}{T} 
        \psi(t - t_p) \psi(t + \tau - t_q) \psi(t + \tau' - t_q) \nonumber \\
    % p, q = r
    & \qquad + \mean{A}^3 \sum \limits_{p = 1}^{K} \sum \limits_{q = 1}^{K} \sum \limits_{r = 1}^{K}  \int \limits_{0}^{T} \frac{\mathrm{d}t_p}{T}  \int \limits_{0}^{T} \frac{\mathrm{d}t_q}{T}  \int \limits_{0}^{T} \frac{\mathrm{d}t_r}{T}  
        \psi(t - t_p) \psi(t + \tau - t_q) \psi(t + \tau' - t_r) .
\end{align}
The sum over the product of the individual pulses is grouped into six sums.
The first sum contains factors with equal pulse arrival times and consists of $K$ terms.
The next three groups contain terms where two pulses occur at the same arrival time,
each group counting $K (K-1)$ terms. The last sum contains the remaining $K (K-1)(K-2)$
terms of the terms where all three pulses occur at different pulse arrival times.

The sum occurring in the four point correlation function may be grouped by equal pulse
arrival time as well. In the latter case, the sum may be split up into group of terms 
where four, three and two pulse arrival times are equal, and in a sum over the remaining 
terms. The sums in each group have $K$, $K(K-1)$, $K(K-1)(K-2)$, and $K(K-1)(K-2)(K-3)$
terms respectively. 

Similar to \Eqnref{integral_1pulse}, we evaluate the integral of the product of
three pulse shapes while neglecting boundary terms to be
\begin{align}
    \int \limits_{0}^{T} \mathrm{d}t_p P_t(t_p) & \psi(t - t_p) \psi(t + \tau - t_p) \psi(t + \tau' - t_p) \nonumber \\
    & \simeq \frac{\taud}{3} \bigexp{\frac{\tau + \tau'}{\taud}} \bigexp{-3 \frac{\max\left(0, \tau, \tau' \right)} {\taud} } 
\end{align}
while the integral of the product of four pulse shapes is given by
\begin{align}
    \int \limits_{0}^{T} \mathrm{d}t_p P_t(t_p) & \psi(t - t_p) \psi(t + \tau - t_p) \psi(t + \tau' - t_p) \psi(t + \tau'' - t_p) \nonumber \\
    & \simeq \frac{\taud}{4} \bigexp{\frac{\tau + \tau' + \tau''}{\taud}} \bigexp{-4 \frac{\max\left(0, \tau, \tau', \tau'' \right)} {\taud} } .
\end{align}
To obtain an expression for the third- and fourth-order correlation functions, these 
integrals are inserted into the correlation function and the resulting expression is
averaged over the total number of pulses. We point out that the $K$ pulses occurring 
in the time interval $[0:T]$ is Poisson distributed and that for a Poisson distributed 
random variable $K$, 
\begin{align*}
    \left \langle \prod\limits_{n=0}^{z} K-n \right \rangle = K^z
\end{align*}
holds. Using this with $Z=2$, the three-point correlation function evaluates to
\begin{align}
    \mean{\Phi(t) \Phi(t + \tau) \Phi(t + \tau')} 
    & = \mean{A}^2 \left[ 2 \frac{\taud}{\tauw} \bigexp{ \frac{\tau + \tau'}{\taud} - 3 \frac{\max(0, \tau, \tau')}{\taud}} \right. \nonumber \\
    & \left. + \left( \left(\frac{\taud}{\tauw}\right)^2 + 1 \right) \bigexp{ \frac{\tau}{\taud} - 2 \frac{\max(0, \tau)}{\taud}} + \left(\frac{\taud}{\tauw}\right)^3 \right].
    \label{eq:corr3-cont}
\end{align}
The four-point correlation function is evaluated the same way. 

To evaluate summations over higher-order correlation function, we note that \Eqnref{corr3-cont}
evaluated at discrete times can be written as
\begin{align}
    \mean{\Phi(t_i) \Phi(t_j) \Phi(t_k)} 
    & = \mean{A}^2 \left[ 2 \left( \frac{\taud}{\tauw} \right) \exp\Bigl( \alpha(2i -j - k) - 3 \alpha \max(0, i-j,j-k)\Bigr) \right. \nonumber \\
    & \left. + \left( \left(\frac{\taud}{\tauw}\right)^2 + 1 \right) \exp \Bigl( \alpha (i-j) - \max(0,i-j) \Bigr)+ \left(\frac{\taud}{\tauw}\right)^3 \right],
    \label{eq:corr3-disc}
\end{align}
where $\tau = \tau_{ij} = \triangle_t \left(i - j\right)$ and
$\tau' = \tau_{jk} = \triangle_t \left( j - k \right)$.
The summations over higher-order correlation functions in \Eqnref{cov_sigmamu_start} 
and \Eqnref{var_sigma2_start} may then be evaluated by approximating the sums by an 
integral, assuming $N \gg 1$, and dividing the integration domain into sectors where 
$i < j < k$, $i < k < j$, $\ldots$. In each of these sectors, the $\max$-functions in 
\Eqnref{corr3-disc} are secular valued so that the integral is well defined. Denoting 
all permutations of the tuple $(i, j, k)$ as $\mathcal{P}_3$, and the respective 
elements of a permutated tuple as $\pi_1$, $\pi_2$, $\pi_3$, we thus have
\begin{align*}
    \sum\limits_{i,j,k=1}^{N} \mean{\Phi(t_i) \Phi(t_j) \Phi(t_k)} & \simeq \int \limits_{0}^{N} \mathrm{d}i\, \mathrm{d}j\, \mathrm{d}k\, 
    \mean{\Phi(t_i) \Phi(t_j) \Phi(t_k)} \times \left(\sum\limits_{\pi \in \mathcal{P}_3} \Theta(\pi_1 - \pi_2) \Theta(\pi_2 - \pi_3) \right) \\
    \sum\limits_{i,j,k,l=1}^{N} \mean{\Phi(t_i) \Phi(t_j) \Phi(t_k) \Phi(t_l)} & \simeq \int \limits_{0}^{N} \mathrm{d}i\, \mathrm{d}j\, \mathrm{d}k\, \mathrm{d}l\,
    \mean{\Phi(t_i) \Phi(t_j) \Phi(t_k) \Phi(t_l)} \times \\ 
    & \left(\sum\limits_{\pi \in \mathcal{P}_4} \Theta(\pi_1 - \pi_2) \Theta(\pi_2 - \pi_3) \Theta(\pi_3 - \pi_4) \right). \\
\end{align*}
These integral are readily evaluated. Inserting them into \Eqnref{cov_sigmamu_start}, 
and  \Eqnref{var_sigma2_start}, yields the expression \Eqnref{COV_mu_var} and \Eqnref{MSE_var}.

%%%%%%%%%%%%%%%%%%%%%%%%%%%%%%%%%%%%%%%%%%%%%%%%%%%%%%%%%%%%%%%%%%%%%%%%%%%%%%%%%%%%%%%%%%%%%%%%%%%%%%%%%%%%
%
% Bibliography
%
%%%%%%%%%%%%%%%%%%%%%%%%%%%%%%%%%%%%%%%%%%%%%%%%%%%%%%%%%%%%%%%%%%%%%%%%%%%%%%%%%%%%%%%%%%%%%%%%%%%%%%%%%%%%

%%%%%%%%%%%%%%%%%%%%%%%%%%%%%%%%%%%%%%%%%%%%%%%%%%%%%%%%%%%%%%%%%%%%%%%%%%%%%%%%%%%%%%%%%%%%%%%%%%%%%%%%%%%%
%
% Figures, synthetic time series. Gamma = 0.1
%
%%%%%%%%%%%%%%%%%%%%%%%%%%%%%%%%%%%%%%%%%%%%%%%%%%%%%%%%%%%%%%%%%%%%%%%%%%%%%%%%%%%%%%%%%%%%%%%%%%%%%%%%%%%%
\newpage
\clearpage
\centering

\newpage

\newpage
\clearpage
\centering

\newpage
\centering

\newpage
\centering

\newpage
\centering
%
%%%%%%%%%%%%%%%%%%%%%%%%%%%%%%%%%%%%%%%%%%%%%%%%%%%%%%%%%%%%%%%%%%%%%%%%%%%%%%%%%%%%%%%%%%%%%%%%%%%%%%%%%%%%
%
% Figures, synthetic time series. Gamma = 10.0
%
%%%%%%%%%%%%%%%%%%%%%%%%%%%%%%%%%%%%%%%%%%%%%%%%%%%%%%%%%%%%%%%%%%%%%%%%%%%%%%%%%%%%%%%%%%%%%%%%%%%%%%%%%%%%
%

\newpage
\clearpage
\centering

\newpage
\clearpage
\centering

\newpage
\clearpage
\centering

\newpage
\centering


\begin{thebibliography}{1}
\bibitem{foot-1} The expected value of the sample coefficient of kurtosis for a sample drawn from
a normal distribution is three. The sample coefficient of excess kurtosis
is found by subtracting 3 from the sample coefficient of kurtosis, see \Eqnref{est_SKmom}
\end{thebibliography}
\end{document}